\documentclass[aps,preprint]{revtex4-1}%
\usepackage{bm}
\usepackage{amsfonts}
\usepackage{amsmath}
\usepackage{amssymb}
\usepackage{graphicx}%
\begin{document}

\title{Neutrino emissivity of $^{3}P_{2}$-$^{3}F_{2}$ superfluid cores in neutron stars}
\author{L. B. Leinson}

\affiliation{Institute of Terrestrial Magnetism, Ionosphere and Radio Wave Propagation RAS,
142190 Troitsk, Moscow Region, Russia}

\begin{abstract}
The influence of the admixture of the $^{3}F_{2}$ state onto collective spin
oscillations and neutrino emission processes in the triplet superfluid neutron
liquid is studied in the BCS approximation. The eigen mode of spin
oscillations with $\omega\simeq\sqrt{\allowbreak58/35}\Delta$ is predicted to
exist in the triplet superfluid neutron condensate besides the already known
mode $\omega\simeq\Delta/\sqrt{5}$. \ Excitation of the high-frequency spin
oscillations in the condensate occurs through the tensor interactions between quasiparticles.

Neutrino energy losses through neutral weak currents are found to consist of
three separate contributions caused by a recombination of broken Cooper pairs
and by weak decays of the collective modes of spin oscillations. Neutrino
decays of the low-frequency spin waves can play an important role in the
cooling scenario of neutron stars. Weak decays of the high-frequency
oscillations that occur only if the tensor forces are taken into account in
the pairing interactions does not modify substantially the total energy
losses. Simple expressions are suggested for the total neutrino emissivity.

\end{abstract}
\maketitle
\startpage{1}

\section{Introduction}

It is considered well established that the inner core of neutron stars
contains a condensate of superfluid neutrons below the critical condensation
temperature $T_{c}$. A superfluid energy gap $\Delta$ arising in the
quasiparticle spectrum suppresses most of the neutrino emission mechanisms in
the volume of the star \cite{YL}, especially when the temperature falls
substantially below the critical value. In this case the number of broken
Cooper pairs rapidly decreases and leads to a strong quenching of the neutrino
emission caused by pair breaking and formation (PBF) processes, which is
considered as the most efficient cooling mechanism of superfluid neutron cores
\cite{Page04,Page09,SY,PPLS}. According to this scenario, at temperatures
$T\simeq0.1T_{c}$, the neutron star enters the epoch of a surface cooling.

The neutron superfluidity in the inner core of neutron stars is believed to
arise owing to pairing of fermions into a triplet state. It is natural to
expect the existence of low-frequency collective modes associated with spin
fluctuations of such a condensate. Previously spin modes have been thoroughly
studied in the $p$-wave superfluid liquid $^{3}He$ with a central interaction
between quasiparticles \cite{Maki,C1,W,Wolfle} . These results cannot be
applied without revision to the triplet superfluid condensate of neutrons,
where the pairing occurs mostly owing to a short-range negative spin-orbit
force of the interaction in the channel of $j=2$.

In a series of papers \cite{L09a,L10a,L10b} we have investigated the
collective spin oscillations in the $^{3}P_{2}$ superfluid neutron liquid
which can be formed because of the strong attractive spin-orbit interaction
between neutrons at high densities. Spin waves with the excitation energy
$\omega=\Delta/\sqrt{5}$ were predicted to exist in such a superfluid
condensate and it has been shown that the spin-wave decay (SWD) through
neutral weak currents leads to a substantial neutrino emission at the lowest
temperatures $T\ll T_{c}$, when all other mechanisms of the neutrino energy
losses are killed by the superfluidity.

In this paper the problem is considered for the case of $^{3}P_{2}-$%
\noindent$^{3}F_{2}$ pairing. We consider the spin eigenmodes of the
superfluid condensate in the case of pairing owing to spin-orbit and tensor
forces. The neutrino emission owing to PBF and SWD processes is calculated.
The calculations are made within the BCS approximation by assuming a
projection of the total angular momentum of the bound pairs $m_{j}=0$ as the
preferable one at supernuclear densities.

The paper is organized as follows. Section II contains some preliminary notes
and outlines some of the important properties of the Green's functions and the
one-loop integrals used below. We recollect the gap equations for the case of
spin-orbit and tensor pairing forces. In Sec. III we discuss the
renormalizations which transform the standard gap equations to a very simple
form valid near the Fermi surface. In Sec. IV we derive, in the BCS
approximation, the equations for anomalous three-point vertices responsible
for the interaction of the neutron superfluid liquid with an external
axial-vector field. In Sec. V we apply the angle average approximation to make
the equations solvable analytically. In Sec. VI we analyze the poles of
anomalous vertices to derive the dispersion of spin-density oscillations in
the condensate. In Sec. VII we calculate the linear response of the superfluid
neutron liquid onto an external axial-vector field. In Sec. VIII we derive
neutrino losses caused by the recombination of broken Cooper pairs and by the
decay of spin waves. In Sec. IX, we evaluate neutrino energy losses in the
$^{3}P_{2}-$\noindent$^{3}F_{2}$ superfluid neutron liquid. Section X contains
a short summary of our findings and the conclusion.

Throughout this paper, we use the standard model of weak interactions, the
system of units $\hbar=c=1$ and the Boltzmann constant $k_{B}=1$.

\section{General approach and notation}

\subsection{Green functions and loop integrals}

The order parameter, $\hat{D}\equiv D_{\alpha\beta}$, arising due to triplet
pairing of quasiparticles, represents a $2\times2$ symmetric matrix in spin
space, $\left(  \alpha,\beta=\uparrow,\downarrow\right)  $. The spin-orbit
interaction among quasiparticles is known to dominate in the nucleon matter of
a high density. Therefore it is conventional to represent the triplet order
parameter of the system $\hat{D}=\sum_{jlm_{j}}\Delta_{jlm_{j}}\Phi
_{\alpha\beta}^{\left(  jlm_{j}\right)  }$ as a superposition of standard
spin-angle functions of the total angular momentum $\left(  j,m_{j}\right)
$,
\begin{equation}
\left(  \Phi_{jlm_{j}}\left(  \mathbf{n}\right)  \right)  _{\alpha\beta}%
\equiv\sum_{m_{s}+m_{l}=m_{j}}\left(  \frac{1}{2}\frac{1}{2}\alpha\beta
|sm_{s}\right)  \left(  slm_{s}m_{l}|jm_{j}\right)  Y_{l,m_{l}}\left(
\mathbf{n}\right)  . \label{sa}%
\end{equation}
The angular dependence of the order parameter is represented by the unit
vector $\mathbf{n=p}/p$ which defines the polar angles $\left(  \theta
,\varphi\right)  $ on the Fermi surface.

For our calculations it is more convenient to use vector notation that
involves a set of mutually orthogonal complex\ vectors $\mathbf{b}_{jlm_{j}%
}\left(  \mathbf{n}\right)  $ defined as%
\begin{equation}
\mathbf{b}_{jlm_{j}}\left(  \mathbf{n}\right)  =-\frac{1}{2}\mathrm{Tr}\left(
\hat{g}\bm{\hat{\sigma}}\hat{\Phi}_{jlm_{j}}\right)  ~,~\mathbf{b}_{jl,-m_{j}%
}=\left(  -\right)  ^{m_{j}}\mathbf{b}_{jlm_{j}}^{\ast}, \label{blm}%
\end{equation}
where $\bm{\hat{\sigma}}=\left(  \hat{\sigma}_{1},\hat{\sigma}_{2},\hat
{\sigma}_{3}\right)  $ are Pauli spin matrices, and $\hat{g}=i\hat{\sigma}%
_{2}$. The vectors $\mathbf{b}_{jlm_{j}}$ obey the normalization condition%
\begin{equation}
\int\frac{d\mathbf{n}}{4\pi}\mathbf{b}_{j^{\prime}l^{\prime}m_{j}^{\prime}%
}^{\ast}\mathbf{b}_{jlm_{j}}=\delta_{jj^{\prime}}\delta_{ll^{\prime}}%
\delta_{m_{j}m_{j}^{\prime}}. \label{lmnorm}%
\end{equation}

Using the vector notation the order parameter is $\hat{D}\left(
\mathbf{n}\right)  =\Delta\mathbf{\bar{b}}\bm{\hat{\sigma}}\hat{g}$, where the
vector $\mathbf{\bar{b}}$ in spin space is defined by the relation%
\begin{equation}
\Delta\left(  p\right)  \mathbf{\bar{b}}\left(  \mathbf{n}\right)
=\sum_{jlm_{j}}\Delta_{jlm_{j}}\left(  p\right)  \mathbf{b}_{jlm_{j}}\left(
\mathbf{n}\right)  . \label{dbbar}%
\end{equation}
Because the ground state order parameter is to be a unitary triplet
\cite{Tamagaki,Takatsuka}, $\mathbf{\bar{b}}\left(  \mathbf{n}\right)  $ is a
real vector which we normalize by the condition
\begin{equation}
\int\frac{d\mathbf{n}}{4\pi}\bar{b}^{2}\left(  \mathbf{n}\right)  =1.
\label{Norm}%
\end{equation}

Making use of the adopted graphical notation for the ordinary and anomalous
propagators, $\hat{G}=\parbox{1cm}{\includegraphics[width=1cm]{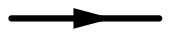}}$,
$\hat{G}^{-}(p)=\parbox{1cm}{\includegraphics[width=1cm,angle=180]{Gn.eps}}$,
$\hat{F}^{(1)}=\parbox{1cm}{\includegraphics[width=1cm]{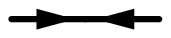}}$\thinspace,
and $\hat{F}^{(2)}=\parbox{1cm}{\includegraphics[width=1cm]{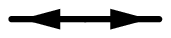}}$%
\thinspace, it is convenient to employ the Matsubara calculation technique for
the system in thermal equilibrium. Then the analytic form of the propagators
is as follows \cite{AGD,Migdal}%
\begin{align}
\hat{G}\left(  p_{\eta},\mathbf{p}\right)   &  =G\left(  p_{\eta}%
,\mathbf{p}\right)  \delta_{\alpha\beta},\ \ \ \ \ \hat{G}^{-}\left(  p_{\eta
},\mathbf{p}\right)  =G^{-}\left(  p_{\eta},\mathbf{p}\right)  \delta
_{\alpha\beta},\nonumber\\
\hat{F}^{\left(  1\right)  }\left(  p_{\eta},\mathbf{p}\right)   &  =F\left(
p_{\eta},\mathbf{p}\right)  \mathbf{\bar{b}}\bm{\hat{\sigma}}\hat
{g},\ \ \ \hat{F}^{\left(  2\right)  }\left(  p_{\eta},\mathbf{p}\right)
=F\left(  p_{\eta},\mathbf{p}\right)  \hat{g}\bm{\hat{\sigma}}\mathbf{\bar{b}%
}, \label{GF}%
\end{align}
where the scalar Green's functions are of the form $G^{-}\left(  p_{\eta
},\mathbf{p}\right)  =G\left(  -p_{\eta},-\mathbf{p}\right)  $ and%
\begin{equation}
G\left(  p_{\eta},\mathbf{p}\right)  =\frac{-ip_{\eta}-\varepsilon
_{\mathbf{p}}}{p_{\eta}^{2}+E_{\mathbf{p}}^{2}},\ F\left(  p_{\eta}%
,\mathbf{p}\right)  =\frac{-\Delta}{p_{\eta}^{2}+E_{\mathbf{p}}^{2}}.
\label{GFc}%
\end{equation}
Here $p_{\eta}\equiv i\pi\left(  2\eta+1\right)  T$ with $\eta=0,\pm1,\pm2...$
is the Matsubara's fermion frequency, and the quasiparticle energy is given
by
\begin{equation}
E_{\mathbf{p}}^{2}=\varepsilon_{p}^{2}+\Delta^{2}\bar{b}^{2}\left(
\mathbf{n}\right)  , \label{Ep}%
\end{equation}
where $\varepsilon_{p}$ is the single-particle spectrum of the normal Fermi
liquid. Near the Fermi surface one has%
\begin{equation}
\varepsilon_{p}=\frac{p^{2}}{2M^{\ast}}-\frac{p_{F}^{2}}{2M^{\ast}}%
\simeq\upsilon_{F}\left(  p-p_{F}\right)  . \label{epsp}%
\end{equation}
The effective mass of a neutron quasiparticle is defined as $M^{\ast}%
=p_{F}/\upsilon_{F}$, where $\upsilon_{F}\ll1$ is the Fermi velocity of the
nonrelativistic neutrons. In the absence of external fields, the gap amplitude
$\Delta\left(  T\right)  $ is real.

The following notation is used below. We denote as $L_{X,X}\left(
\omega,\mathbf{q;p}\right)  $ the analytical continuation of the Matsubara
sums:%
\begin{equation}
L_{XX^{\prime}}\left(  \omega_{\kappa},\mathbf{p+}\frac{\mathbf{q}}%
{2}\mathbf{;p-}\frac{\mathbf{q}}{2}\right)  =T\sum_{\eta}X\left(  p_{\eta
}+\omega_{\kappa},\mathbf{p+}\frac{\mathbf{q}}{2}\right)  X^{\prime}\left(
p_{\eta},\mathbf{p-}\frac{\mathbf{q}}{2}\right)  , \label{LXX}%
\end{equation}
where $X,X^{\prime}\in G,F,G^{-}$, and $\omega_{\kappa}=2i\pi\kappa T$ with
$\kappa=0,\pm1,\pm2...$. We divide the integration over the momentum space
into integration over the solid angle and over the energy according to%
\begin{equation}
\int\frac{d^{3}p}{\left(  2\pi\right)  ^{3}}...=\varrho\int\frac{d\mathbf{n}%
}{4\pi}\frac{1}{2}\int_{-\infty}^{\infty}d\varepsilon_{\mathbf{p}}...,
\label{1}%
\end{equation}
where $\varrho=p_{F}M^{\ast}/\pi^{2}$ is the density of states near the Fermi
surface in the normal state, and operate with integrals over the quasiparticle
energy:%
\begin{equation}
\mathcal{I}_{XX^{\prime}}\left(  \omega,\mathbf{n,q};T\right)  \equiv\frac
{1}{2}\int_{-\infty}^{\infty}d\varepsilon_{p}L_{XX^{\prime}}\left(
\omega,\mathbf{p+}\frac{\mathbf{q}}{2}\mathbf{,p-}\frac{\mathbf{q}}{2}\right)
. \label{IXX}%
\end{equation}
These are functions of $\omega$, $\mathbf{q}$ and the direction of a
quasiparticle momentum $\mathbf{n}$.

The loop integrals (\ref{IXX}) possess the following properties which can be
verified by a straightforward calculation:%
\begin{equation}
\mathcal{I}_{G^{-}G}=\mathcal{I}_{GG^{-}}~,~\mathcal{I}_{GF}=-\mathcal{I}%
_{FG}~,~\mathcal{I}_{G^{-}F}=-\mathcal{I}_{FG^{-}}, \label{Leg}%
\end{equation}%
\begin{equation}
\mathcal{I}_{G^{-}F}+\mathcal{I}_{FG}=\frac{\omega}{\Delta}\mathcal{I}_{FF},
\label{gf}%
\end{equation}%
\begin{equation}
\mathcal{I}_{G^{-}F}-\mathcal{I}_{FG}=-\frac{\mathbf{qv}}{\Delta}%
\mathcal{I}_{FF}. \label{ff1}%
\end{equation}
For arbitrary $\omega,\mathbf{q},T$ one can obtain also
\begin{equation}
\mathcal{I}_{GG^{-}}+\bar{b}^{2}\mathcal{I}_{FF}=A+\frac{\omega^{2}-\left(
\mathbf{qv}\right)  ^{2}}{2\Delta^{2}}\mathcal{I}_{FF}, \label{FF}%
\end{equation}
where $\mathbf{v}$ is a vector with the magnitude of the Fermi velocity
$\upsilon_{F}$ and the direction of $\mathbf{n}$, and%
\begin{equation}
A\left(  \mathbf{n}\right)  \equiv\left[  \mathcal{I}_{GG^{-}}\left(
\mathbf{n}\right)  +\bar{b}^{2}\left(  \mathbf{n}\right)  \mathcal{I}%
_{FF}\left(  \mathbf{n}\right)  \right]  _{\omega=0,\mathbf{q}=0}. \label{A}%
\end{equation}

\subsection{Gap equation}

The block of the interaction diagrams irreducible in the channel of two
quasiparticles, $\Gamma_{\alpha\beta,\gamma\delta}$, is usually generated by
the expansion over spin-angle functions (\ref{sa}). The spin-orbit interaction
among quasiparticles is known to dominate at high densities. In this case the
spin $\mathbf{s}$ and orbital momentum $\mathbf{l}$ of the pair cease to be
conserved separately. Thus the complete list of channels participating in the
triplet-spin $P$-wave pairing includes the pair states with $j=0,1,2$, and
$\left\vert m_{j}\right\vert \leq j$. The pairing occurs in the state with
$j=2$ because the attractive interaction in this channel strongly dominates.
The tensor components of the neutron-neutron interaction are known also to
exert some influence on pair formation in dense neutron matter, favoring the
condensation of pairs in the $^{3}P_{2}+$\noindent$^{3}F_{2}$ state, but the
contributions from $^{3}P_{2}\rightarrow$\noindent$^{3}P_{0}$ or $^{3}%
P_{2}\rightarrow$\noindent$^{3}P_{1}$ transitions are deemed to be
unimportant. Hence we take the approximation to neglect the $j=0,1$ coupling
throughout this paper. From now on we omit the suffix $j$ everywhere by
assuming that the pairing occurs into the state with $j=2$. Then, in the
vector notation, the pairing interaction is of the form%

\begin{equation}
\varrho\Gamma_{\alpha\beta,\gamma\delta}\left(  \mathbf{p,p}^{\prime}\right)
=\sum_{l^{\prime}lm_{j}}\left(  -1\right)  ^{\frac{l-l^{\prime}}{2}%
}\mathcal{V}_{ll^{\prime}}\left(  p,p^{\prime}\right)  \left(  \mathbf{b}%
_{lm_{j}}(\mathbf{n})\bm{\hat{\sigma}}\hat{g}\right)  _{\alpha\beta}\left(
\hat{g}\bm{\hat{\sigma}}\mathbf{b}_{l^{\prime}m_{j}}^{\ast}(\mathbf{n}%
^{\prime})\right)  _{\gamma\delta}, \label{ppint}%
\end{equation}
where $\allowbreak$the pairing matrix elements $\mathcal{V}_{ll^{\prime}%
}\left(  p,p^{\prime}\right)  $ with $l,l^{\prime}=j\pm1=1,3$ are the
corresponding interaction amplitudes.

The ground-state problem is normally treated in terms of the set of equations
for the coupled partial-wave amplitudes $\Delta_{lm_{j}}$
\cite{Amundsen,Khod,Baldo,Elg,Khodel,Schwenk}. Making use of the identity%
\begin{equation}
{\frac{1}{2E_{\mathbf{p}}}}\tanh{\frac{E_{\mathbf{p}}}{2T}\equiv T\sum_{\eta
}\frac{1}{p_{\eta}^{2}+E_{\mathbf{p}}^{2}}},\label{sum}%
\end{equation}
one can obtain the standard set of equations for the triplet partial
amplitudes $\Delta_{lm_{j}}$ in the form%
\begin{align}
\Delta_{lm_{j}}\left(  p\right)   &  =-\sum_{l^{\prime}=1,3}\frac{1}{2\varrho
}\int dp^{\prime}p^{\prime2}\left(  -1\right)  ^{\frac{l-l^{\prime}}{2}%
}\mathcal{V}_{ll^{\prime}}\left(  p,p^{\prime}\right)  \nonumber\\
&  \times\Delta\left(  p^{\prime}\right)  \left\langle \mathbf{b}_{l^{\prime
}m_{j}}^{\ast}(\mathbf{n}^{\prime})\mathbf{\bar{b}}(\mathbf{n}^{\prime}%
)T\sum_{\eta}\frac{1}{p_{\eta}^{2}+E_{\mathbf{p}^{\prime}}^{2}}\right\rangle
,\label{gap}%
\end{align}
Here and in what follows we use the angle brackets to denote angle averages,
\[
\left\langle ...\right\rangle \equiv\frac{1}{4\pi}\int d\mathbf{n}....
\]
Notice that%
\begin{equation}
\frac{1}{p_{\eta}^{2}+E_{\mathbf{p}}^{2}}\equiv G\left(  p_{\eta}%
,\mathbf{p}\right)  G^{-}\left(  p_{\eta},\mathbf{p}\right)  +\bar{b}%
^{2}F\left(  p_{\eta},\mathbf{p}\right)  F\left(  p_{\eta},\mathbf{p}\right)
,\label{GGm}%
\end{equation}
and the gap equation (\ref{gap}) can be identically written as%
\begin{align}
\Delta_{lm_{j}}\left(  p\right)   &  =-\sum_{l^{\prime}}\frac{1}{2\varrho}\int
dp^{\prime}p^{\prime2}\left(  -1\right)  ^{\frac{l-l^{\prime}}{2}}%
\mathcal{V}_{ll^{\prime}}\left(  p,p^{\prime}\right)  \Delta\left(  p^{\prime
}\right)  \nonumber\\
&  \times\left\langle \mathbf{b}_{l^{\prime}m_{j}}^{\ast}(\mathbf{n}^{\prime
})\mathbf{\bar{b}}(\mathbf{n}^{\prime})\left[  L_{GG^{-}}+\bar{b}^{2}%
L_{FF}\right]  _{\omega=0,\mathbf{q}=0}\right\rangle .\label{gapL}%
\end{align}

\section{Renormalizations}

Both the gap equation (\ref{gapL}) and the vertex equation (\ref{Beq}) involve
integrations over the regions far from the Fermi surface while we are
interested in the processes occuring in a vicinity of the Fermi sphere. To get
rid of the integration over the far regions we renormalize the interaction as
suggested in Refs. \cite{Leggett,Leggett1}: We define
\begin{align*}
\mathsf{V}_{ll^{\prime}}\left(  p,p^{\prime}\right)   &  =\mathcal{V}%
_{ll^{\prime}}\left(  p,p^{\prime}\right)  -\sum_{l^{\prime\prime}}\int
\frac{dp^{\prime\prime}p^{\prime\prime2}}{2\pi^{2}}\mathcal{V}_{ll^{\prime
\prime}}\left(  p,p^{\prime\prime}\right)  L_{GG^{-}}^{\left(  N\right)
}\left(  p^{\prime\prime}\right)  \mathsf{V}_{l^{\prime\prime}l^{\prime}%
}\left(  p^{\prime\prime},p^{\prime}\right)  \\
&  =\mathcal{V}_{ll^{\prime}}\left(  p,p^{\prime}\right)  -\sum_{l^{\prime
\prime}}\int\frac{dp^{\prime\prime}p^{\prime\prime2}}{2\pi^{2}}\mathsf{V}%
_{ll^{\prime\prime}}\left(  p,p^{\prime\prime}\right)  L_{GG^{-}}^{\left(
N\right)  }\left(  p^{\prime\prime}\right)  \mathcal{V}_{l^{\prime\prime
}l^{\prime}}\left(  p^{\prime\prime},p^{\prime}\right)  ,
\end{align*}
where the loop$~L_{GG^{-}}^{\left(  N\right)  }\left(  p^{\prime\prime
}\right)  $ is evaluated in the normal (nonsuperfluid) state. Using the
identity%
\[
\frac{\Delta\left(  p\right)  }{\Delta_{lm_{j}}\left(  p\right)  }\int
\frac{d\mathbf{n}}{4\pi}\left(  \mathbf{b}_{lm_{j}}^{\ast}\mathbf{\bar{b}%
}\right)  \equiv1
\]
one can recast the above as%
\begin{align*}
\mathsf{V}_{ll^{\prime}}\left(  p,p^{\prime}\right)   &  =\mathcal{V}%
_{ll^{\prime}}\left(  p,p^{\prime}\right)  -\sum_{l^{\prime\prime}}\int
\frac{d^{3}\mathbf{p}^{\prime\prime}}{8\pi^{3}}\mathsf{V}_{ll^{\prime}}\left(
p,p^{\prime\prime}\right)  \\
&  \times\frac{\Delta\left(  p^{\prime\prime}\right)  }{\Delta_{l^{\prime
\prime}m_{j}}\left(  p^{\prime\prime}\right)  }\left(  \mathbf{b}%
_{l^{\prime\prime}m_{j}}^{\ast}\mathbf{\bar{b}}\right)  _{\mathbf{n}%
^{\prime\prime}}L_{GG^{-}}^{\left(  N\right)  }\left(  p^{\prime\prime
}\right)  \mathcal{V}_{l^{\prime\prime}l^{\prime}}\left(  p^{\prime\prime
},p^{\prime}\right)  .
\end{align*}

Then it can be shown \cite{L10a} that we may everywhere substitute
$\mathsf{V}_{ll^{\prime}}$ for $\mathcal{V}_{ll^{\prime}}$ provided that at
the same time, we understand by the $L_{GG^{-}}$ element, the subtracted
quantity $L_{GG^{-}}$\noindent$-$\noindent$L_{GG^{-}}^{\left(  N\right)
}\left(  p^{\prime\prime}\right)  $ [$L_{GG^{-}}^{\left(  N\right)  }$ is to
be evaluated for $\omega=0,\mathbf{q}=0$ in all cases]. The gap equation
(\ref{gapL}) becomes of the form%
\begin{equation}
\Delta_{lm_{j}}=-\sum_{l^{\prime}}i^{l-l^{\prime}}\mathsf{V}_{ll^{\prime}%
}\Delta\left\langle \mathbf{b}_{l^{\prime}m_{j}}^{\ast}(\mathbf{n}%
)\mathbf{\bar{b}}(\mathbf{n})A\left(  \mathbf{n}\right)  \right\rangle
~\label{GAP}%
\end{equation}
which is valid in the narrow vicinity of the Fermi surface, where the smooth
functions $\Delta_{lm_{j}}\left(  p\right)  $, $\mathsf{V}_{ll^{\prime}%
}\left(  p,p^{\prime}\right)  $, and $\Delta\left(  p\right)  $ may be
replaced with constants $\Delta\left(  p\right)  \simeq\Delta\left(
p_{F}\right)  \equiv\Delta$, etc. The function (\ref{A}) is now to be
understood as%
\begin{equation}
A\left(  \mathbf{n}\right)  \rightarrow\left[  \mathcal{I}_{GG^{-}%
}-\mathcal{I}_{GG^{-}}^{\left(  N\right)  }+\bar{b}^{2}\mathcal{I}%
_{FF}\right]  _{\omega=0,\mathbf{q}=0}.\label{Ar}%
\end{equation}
It can be found explicitly by performing the Matsubara summation:%
\begin{equation}
A\left(  \mathbf{n}\right)  =\frac{1}{2}\int_{0}^{\infty}d\varepsilon\left(
\frac{1}{\sqrt{\varepsilon^{2}+\Delta^{2}\bar{b}^{2}}}\tanh\frac
{\sqrt{\varepsilon^{2}+\Delta^{2}\bar{b}^{2}}}{2T}-\frac{1}{\varepsilon}%
\tanh\frac{\varepsilon}{2T}\right)  .\label{An}%
\end{equation}
The renormalization of Eq. (\ref{Beq}) also reduces to the replacements
$\mathcal{V}_{ll^{\prime}}\rightarrow\mathsf{V}_{ll^{\prime}}$. The function
$A\left(  \mathbf{n}\right)  $ should be replaced by the expression (\ref{An}).

\section{Vertex equations}

\begin{figure}[h]
\includegraphics{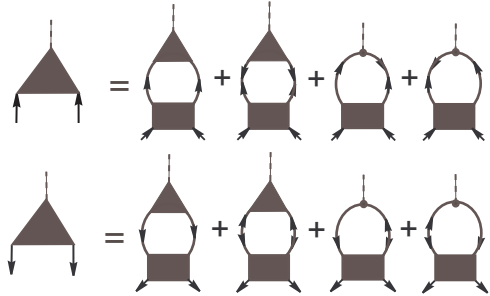}\caption{Dyson's equations for the anomalous
vertices. The ordinary vertices are shown by small filled circles. The shaded
rectangle represents the pairing interaction.}%
\label{fig1}%
\end{figure}

We are interested in the linear medium response onto an external axial-vector
field. The field interaction with a superfluid should be described with the
aid of two ordinary and two anomalous three-point effective vertices. In the
BCS approximation, the ordinary axial-vector vertices of a nonrelativistic
particle and a hole are to be taken as $\bm{\hat{\sigma}}$ and
$\bm{\hat{\sigma}}^{T}$, respectively. The anomalous effective vertices,
$\mathbf{\hat{T}}^{\left(  1\right)  }\left(  \mathbf{n;}\omega,\mathbf{q}%
\right)  $ and $\mathbf{\hat{T}}^{\left(  2\right)  }\left(  \mathbf{n;}%
\omega,\mathbf{q}\right)  $ are given by the infinite sums of the diagrams
taking account of the pairing interaction in the ladder approximation
\cite{Larkin}. These $2\times2$ vector matrices are to satisfy the Dyson's
equations symbolically depicted by graphs in Fig. \ref{fig1}.

The analytic form of the diagrams in Fig. 1 is derived in Ref. \cite{L09a}. We
are interested in the neutrino energy losses through neutral weak currents. In
the case of a nonrelativistic medium the relevant input for this calculation
are the effective weak vertices at zero momentum transfer. This substantially
simplifies the problem. After some algebraic manipulations the BCS equations
for anomalous vertices at $\mathbf{q}=0$ can be found in the following form
(we omit for brevity the dependence of functions on $\omega$\textbf{)}:
\begin{gather}
\mathbf{\hat{T}}^{\left(  1\right)  }\left(  \mathbf{n}\right)  =\sum_{lm_{j}%
}\bm{\hat{\sigma}}\mathbf{b}_{lm_{j}}(\mathbf{n})\hat{g}\sum_{l^{\prime}%
}\mathcal{V}_{ll^{\prime}}\frac{1}{2}\left\langle \mathcal{I}_{GG^{-}%
}\mathrm{Tr}\left[  \hat{g}\left(  \bm{\hat{\sigma}}\mathbf{b}_{l^{\prime
}m_{j}}^{\ast}\right)  \mathbf{\hat{T}}^{\left(  1\right)  }\right]  \right.
\nonumber\\
\left.  -\mathcal{I}_{FF}\mathrm{Tr}\left[  \left(
\bm{\hat{\sigma}}\mathbf{b}_{l^{\prime}m_{j}}^{\ast}\right)  \left(
\bm{\hat{\sigma}}\mathbf{\bar{b}}\right)  \hat{g}\mathbf{\hat{T}}^{\left(
2\right)  }\left(  \bm{\hat{\sigma}}\mathbf{\bar{b}}\right)  \right]
-\frac{\omega}{\Delta}\mathcal{I}_{FF}2i\left(  \mathbf{b}_{l^{\prime}m_{j}%
}^{\ast}\mathbf{\times\bar{b}}\right)  \right\rangle , \label{EqT1}%
\end{gather}%
\begin{gather}
\mathbf{\hat{T}}^{\left(  2\right)  }\left(  \mathbf{n}\right)  =\sum_{lm_{j}%
}\hat{g}\bm{\hat{\sigma}}\mathbf{b}_{lm_{j}}^{\ast}(\mathbf{n})\sum
_{l^{\prime}}\mathcal{V}_{ll^{\prime}}\frac{1}{2}\left\langle \mathcal{I}%
_{G^{-}G}\mathrm{Tr}\left[  \left(  \bm{\hat{\sigma}}\mathbf{b}_{l^{\prime
}m_{j}}\right)  \hat{g}\mathbf{\hat{T}}^{\left(  2\right)  }\right]  \right.
\nonumber\\
\left.  -\mathcal{I}_{FF}\mathrm{Tr}\left[  \left(
\bm{\hat{\sigma}}\mathbf{b}_{l^{\prime}m_{j}}\right)  \left(
\bm{\hat{\sigma}}\mathbf{\bar{b}}\right)  \mathbf{\hat{T}}^{\left(  1\right)
}\hat{g}\left(  \bm{\hat{\sigma}}\mathbf{\bar{b}}\right)  \right]
-\frac{\omega}{\Delta}\mathcal{I}_{FF}2i\left(  \mathbf{b}_{l^{\prime}m_{j}%
}\mathbf{\times\bar{b}}\right)  \right\rangle . \label{EqT2}%
\end{gather}
Inspection of the equations reveals that the anomalous axial-vector vertices
can be found in the following form
\begin{equation}
\mathbf{\hat{T}}^{\left(  1\right)  }\left(  \mathbf{n},\omega\right)
=\sum_{lm_{j}}\mathbf{B}_{lm_{j}}^{\left(  1\right)  }\left(  \omega\right)
\left(  \bm{\hat{\sigma}}\mathbf{b}_{lm_{j}}\right)  \hat{g}, \label{T1A}%
\end{equation}%
\begin{equation}
\mathbf{\hat{T}}^{\left(  2\right)  }\left(  \mathbf{n},\omega\right)
=\sum_{lm_{j}}\mathbf{B}_{lm_{j}}^{\left(  2\right)  }\left(  \omega\right)
\hat{g}\left(  \bm{\hat{\sigma}}\mathbf{b}_{lm_{j}}^{\ast}\right)  .
\label{T2A}%
\end{equation}
Insertion of these expressions into Eqs. (\ref{EqT1}) and (\ref{EqT2}) makes
it possible to obtain the following equations%
\begin{gather}
\mathbf{B}_{lm_{j}}^{+}=-\sum_{l^{\prime}}\mathcal{V}_{ll^{\prime}}\left[
\sum_{l^{\prime\prime}m_{j}^{\prime}}\left\langle \left(  \mathcal{I}_{GG^{-}%
}+\bar{b}^{2}\mathcal{I}_{FF}\right)  \left(  \mathbf{b}_{l^{\prime}m_{j}%
}^{\ast}\mathbf{b}_{l^{\prime\prime}m_{j}^{\prime}}\right)  \right\rangle
\mathbf{B}_{l^{\prime\prime}m_{j}^{\prime}}^{+}\right. \nonumber\\
\left.  -2\sum_{l^{\prime\prime}m_{j}^{\prime}}\left\langle \left(
\mathbf{\bar{b}b}_{l^{\prime}m_{j}}^{\ast}\right)  \left(  \mathbf{\bar{b}%
b}_{l^{\prime\prime}m_{j}^{\prime}}\right)  \mathcal{I}_{FF}\right\rangle
\mathbf{B}_{l^{\prime\prime}m_{j}^{\prime}}^{+}+\frac{\omega}{\Delta
}\left\langle \mathcal{I}_{FF}i\left(  \mathbf{b}_{l^{\prime}m_{j}}^{\ast
}\mathbf{\times\bar{b}}\right)  \right\rangle \right]  , \label{Bp}%
\end{gather}%
\begin{align}
\mathbf{B}_{lm_{j}}^{-}  &  =-\sum_{l^{\prime}}\mathcal{V}_{ll^{\prime}}%
\sum_{l^{\prime\prime}m_{j}^{\prime}}\left[  \left\langle \left(
\mathcal{I}_{GG^{-}}-\bar{b}^{2}\mathcal{I}_{FF}\right)  \left(
\mathbf{b}_{l^{\prime}m_{j}}^{\ast}\mathbf{b}_{l^{\prime\prime}m_{j}^{\prime}%
}\right)  \right\rangle \right. \nonumber\\
&  \left.  +2\left\langle \left(  \mathbf{\bar{b}b}_{l^{\prime}m_{j}}^{\ast
}\right)  \left(  \mathbf{\bar{b}b}_{l^{\prime\prime}m_{j}^{\prime}}\right)
\mathcal{I}_{FF}\right\rangle \right]  \mathbf{B}_{l^{\prime\prime}%
m_{j}^{\prime}}^{-}, \label{Bm}%
\end{align}
where the new unknown vector functions are defined as%
\begin{equation}
\mathbf{B}_{lm_{j}}^{\left(  \pm\right)  }=\frac{1}{2}\left[  \mathbf{B}%
_{lm_{j}}^{\left(  1\right)  }\pm\left(  -\right)  ^{m_{j}}\mathbf{B}%
_{l,-m_{j}}^{\left(  2\right)  }\right]  . \label{Bpm}%
\end{equation}
The uniform Eqs. (\ref{Bm}) have nontrivial solutions if the determinant of
the system equals zero. This condition could be considered as the dispersion
equation for the eigen modes of oscillations. However, the solutions obtained
in this way would be spurious, because the physical solution must develop a
pole in the vertex function when the frequency approaches the eigenvalue. In
the case of uniform equations the solution remains finite at the resonant
frequency. Therefore only trivial solutions $\mathbf{B}_{lm_{j}}^{-}=0$ are
physically meaningful. We then obtain%
\begin{equation}
\mathbf{B}_{lm_{j}}^{+}\mathbf{=B}_{lm_{j}}^{\left(  1\right)  }=\left(
-\right)  ^{m_{j}}\mathbf{B}_{l,-m_{j}}^{\left(  2\right)  } \label{BpBm}%
\end{equation}
From now on we omit "plus" in the notation by assuming $\mathbf{B}_{lm_{j}%
}\equiv\mathbf{B}_{lm_{j}}^{+}$.

By making use of Eq. (\ref{FF}) and denoting $\mathcal{I}_{FF}\left(
\mathbf{n},\omega,\mathbf{q}=0\right)  =\mathcal{I}_{0}\left(  \mathbf{n,}%
\omega;T\right)  $, where
\begin{equation}
\mathcal{I}_{0}\left(  \mathbf{n,}\omega\right)  =\int_{0}^{\infty}%
\frac{d\varepsilon_{p}}{E_{\mathbf{p}}}\ \frac{\Delta^{2}}{4E_{\mathbf{p}}%
^{2}-\left(  \omega+i0\right)  ^{2}}\tanh\frac{E_{\mathbf{p}}}{2T},
\label{FFq0}%
\end{equation}
one can obtain Eq. (\ref{Bp}) in the form%
\begin{gather}
\mathbf{B}_{lm_{j}}=-\sum_{l^{\prime}}\mathcal{V}_{ll^{\prime}}\left\{
\sum\nolimits_{l^{\prime\prime}m_{j}^{\prime}}\mathbf{B}_{l^{\prime\prime
}m_{j}^{\prime}}\left\langle A\left(  \mathbf{n}\right)  \left(
\mathbf{b}_{l^{\prime}m_{j}}^{\ast}\mathbf{b}_{l^{\prime\prime}m_{j}^{\prime}%
}\right)  \right\rangle \right. \nonumber\\
+\sum_{l^{\prime\prime}m_{j}^{\prime}}\mathbf{B}_{l^{\prime\prime}%
m_{j}^{\prime}}\left\langle \mathcal{I}_{0}\left(  \mathbf{n}\right)  \left[
\frac{\omega^{2}}{2\Delta^{2}}\left(  \mathbf{b}_{l^{\prime}m_{j}}^{\ast
}\mathbf{b}_{l^{\prime\prime}m_{j}^{\prime}}\right)  -2\left(  \mathbf{\bar
{b}b}_{l^{\prime}m_{j}}^{\ast}\right)  \left(  \mathbf{\bar{b}b}%
_{l^{\prime\prime}m_{j}^{\prime}}\right)  \right]  \right\rangle \nonumber\\
\left.  +i\frac{\omega}{\Delta}\left\langle \mathcal{I}_{0}\left(
\mathbf{n}\right)  \left(  \mathbf{b}_{l^{\prime}m_{j}}^{\ast}\mathbf{\times
\bar{b}}\right)  \right\rangle \right\}  , \label{Beq}%
\end{gather}
This equation is to be solved together with the gap equation (\ref{gapL}).

\section{Angle average approximation}

The angle dependence of the functions $A\left(  \mathbf{n}\right)  $ and
$\mathcal{I}_{FF}\left(  \mathbf{n}\right)  $ arises owing to anisotropy of
the square of the energy gap $\Delta^{2}\left(  p^{\prime}\right)  \bar{b}%
^{2}\left(  \mathbf{n}\right)  $ entering the energy of a quasiparticle. The
consideration can be substantially simplified using the angle average
approximation, that is, replacing the anisotropic energy gap with its angle
average, $\Delta^{2}\bar{b}^{2}\rightarrow\left\langle \Delta^{2}\bar{b}%
^{2}\right\rangle =\Delta^{2}$. In Refs. \cite{TakTam,VVkh,BEEHS}, it has been
shown that the angle average approximation is an excellent approximation to
the true solution, as long as one is only interested in the average value of
the gap near the Fermi surface, and not the angular dependence of the gap
functions. After this replacement the angle integration becomes trivial.
Making use of the orthogonality relations (\ref{lmnorm}) after the
renormalizations from Eq. (\ref{Beq}) we get a set of linear equations for
each value of $m_{j}$%
\begin{align}
\mathbf{B}_{lm_{j}}  &  =-\sum_{l^{\prime}}\mathsf{V}_{ll^{\prime}}\left\{
\mathbf{B}_{l^{\prime}m_{j}}\left(  A_{\mathsf{av}}+\frac{\omega^{2}}%
{2\Delta^{2}}\mathcal{I}_{\mathsf{av}}\right)  \right. \nonumber\\
&  -2\sum_{l^{\prime\prime}m_{j}^{\prime}}\mathbf{B}_{l^{\prime\prime}%
m_{j}^{\prime}}\left\langle \left(  \mathbf{\bar{b}b}_{l^{\prime}m_{j}}^{\ast
}\right)  \left(  \mathbf{\bar{b}b}_{l^{\prime\prime}m_{j}^{\prime}}\right)
\right\rangle \mathcal{I}_{\mathsf{av}}\left.  +i\frac{\omega}{\Delta
}\mathcal{I}_{\mathsf{av}}\left\langle \mathbf{b}_{l^{\prime}m_{j}}^{\ast
}\mathbf{\times\bar{b}}\right\rangle \right\}  , \label{Bj}%
\end{align}
where%
\begin{equation}
A_{\mathsf{av}}=\frac{1}{2}\int_{0}^{\infty}d\varepsilon\left(  \frac{1}%
{E}\tanh\frac{E}{2T}-\frac{1}{\varepsilon}\tanh\frac{\varepsilon}{2T}\right)
, \label{Aav}%
\end{equation}
and%
\begin{equation}
\mathcal{I}_{\mathsf{av}}=\int_{0}^{\infty}\frac{d\varepsilon}{E}%
\ \frac{\Delta^{2}}{4E^{2}-\omega^{2}}\tanh\frac{E}{2T} \label{Iav}%
\end{equation}
with $E=\sqrt{\varepsilon^{2}+\Delta^{2}}$.

The gap equation (\ref{GAP}) becomes of the form
\begin{equation}
\Delta_{lm_{j}}=-\sum_{l^{\prime}}i^{l-l^{\prime}}\mathsf{V}_{ll^{\prime}%
}\Delta_{l^{\prime}m_{j}}A_{\mathsf{av}}. \label{GAPav}%
\end{equation}

It is convenient to write the Eqs. (\ref{Bj}) and (\ref{GAPav}) as two matrix
equations. The corresponding vertex equation is
\begin{gather}
\left(
\begin{array}
[c]{c}%
\mathbf{B}_{1m_{j}}\\
\mathbf{B}_{3,m_{j}}%
\end{array}
\right)  =-\left(
\begin{array}
[c]{cc}%
\mathsf{V}_{11} & -\mathsf{V}_{13}\\
-\mathsf{V}_{13} & \mathsf{V}_{33}%
\end{array}
\right)  \left\{  \left(
\begin{array}
[c]{c}%
\left(  A_{\mathsf{av}}+\frac{\omega^{2}}{2\Delta^{2}}\mathcal{I}%
_{\mathsf{av}}\right)  \mathbf{B}_{1m_{j}}\\
\left(  A_{\mathsf{av}}+\frac{\omega^{2}}{2\Delta^{2}}\mathcal{I}%
_{\mathsf{av}}\right)  \mathbf{B}_{3m_{j}}%
\end{array}
\right)  \right. \nonumber\\
\left.  -\left(
\begin{array}
[c]{c}%
2\mathcal{I}_{\mathsf{av}}\sum_{lm_{j}^{\prime}}\left\langle \left(
\mathbf{\bar{b}b}_{1m_{j}}^{\ast}\right)  \left(  \mathbf{\bar{b}b}%
_{lm_{j}^{\prime}}\right)  \right\rangle \mathbf{B}_{lm_{j}^{\prime}}\\
2\mathcal{I}_{\mathsf{av}}\sum_{lm_{j}^{\prime}}\left\langle \left(
\mathbf{\bar{b}b}_{3m_{j}}^{\ast}\right)  \left(  \mathbf{\bar{b}b}%
_{lm_{j}^{\prime}}\right)  \right\rangle \mathbf{B}_{lm_{j}^{\prime}}%
\end{array}
\right)  +\frac{\omega}{\Delta}i\left(
\begin{array}
[c]{c}%
\mathcal{I}_{\mathsf{av}}\left\langle \mathbf{b}_{1m_{j}}^{\ast}%
\mathbf{\times\bar{b}}\right\rangle \\
\mathcal{I}_{\mathsf{av}}\left\langle \mathbf{b}_{3m_{j}}^{\ast}%
\mathbf{\times\bar{b}}\right\rangle
\end{array}
\right)  \right\}  , \label{Bj2matr}%
\end{gather}
and the gap equation becomes of the form%
\begin{equation}
\left(
\begin{array}
[c]{c}%
\Delta_{1m_{j}}\\
\Delta_{3m_{j}}%
\end{array}
\right)  =-\left(
\begin{array}
[c]{cc}%
\mathsf{V}_{11} & -\mathsf{V}_{13}\\
-\mathsf{V}_{13} & \mathsf{V}_{33}%
\end{array}
\right)  \left(
\begin{array}
[c]{c}%
\Delta_{1m_{j}}A_{\mathsf{av}}\\
\Delta_{3m_{j}}A_{\mathsf{av}}%
\end{array}
\right)  . \label{gapeq}%
\end{equation}
In obtaining the equations the fact is used that the interaction matrix is
symmetric on the Fermi surface, $\mathsf{V}_{31}=\mathsf{V}_{13}$.

The interaction matrix which enters Eqs. (\ref{Bj2matr}) and (\ref{gapeq}) can
be diagonalized by unitary transformations $V^{^{\prime}}=UVU^{\dagger}$ with
$U$ being the unitary matrix
\begin{equation}
U=\left(  U^{-1}\right)  ^{\dagger}=\frac{1}{\left(  V_{+}+V_{-}\right)
^{\frac{1}{2}}\allowbreak}\left(
\begin{array}
[c]{cc}%
\sqrt{V_{+}} & \sqrt{V_{-}}\\
-\sqrt{V_{-}} & \sqrt{V_{+}}%
\end{array}
\right)  , \label{Umatr}%
\end{equation}
where$\allowbreak$ $V_{\pm}=\sqrt{\left(  \mathsf{V}_{33}-\mathsf{V}%
_{11}\right)  ^{2}+4\mathsf{V}_{13}^{2}}\pm\left(  \mathsf{V}_{33}%
-\mathsf{V}_{11}\right)  $.

One has $UVU^{\dagger}=\operatorname*{diag}\left(  W_{-},W_{+}\right)  $ with%
\begin{equation}
W_{\pm}=\frac{1}{2}\left(  \mathsf{V}_{33}+\mathsf{V}_{11}\pm\sqrt{\left(
\mathsf{V}_{33}-\mathsf{V}_{11}\right)  ^{2}+4\mathsf{V}_{13}^{2}}\right)  .
\label{W}%
\end{equation}
Applying the unitary transformation $U$\ to the gap equations (\ref{gapeq})
yields two coupled equations:%
\begin{equation}
\sqrt{V_{+}}\Delta_{1m_{j}}+\sqrt{V_{-}}\Delta_{3m_{j}}=-W_{-}\left(
\sqrt{V_{+}}\Delta_{1m_{j}}+\sqrt{V_{-}}\Delta_{3m_{j}}\right)  A_{\mathsf{av}%
}, \label{Ge1}%
\end{equation}%
\begin{equation}
\sqrt{V_{-}}\Delta_{1m_{j}}-\sqrt{V_{+}}\Delta_{3m_{j}}=-W_{+}\left(
\sqrt{V_{-}}\Delta_{1m_{j}}-\sqrt{V_{+}}\Delta_{3m_{j}}\right)  A_{\mathsf{av}%
}. \label{Ge2}%
\end{equation}
In Eq. (\ref{Bj2matr}), the interaction matrix can be also diagonalized by the
unitary transformation (\ref{Umatr}). Further simplification is possible owing
to the fact that by virtue of Eqs. (\ref{Ge1}) and (\ref{Ge2}), the coupling
constants $W_{\pm}$ can be removed out of the equations. This results in the
set of equations%
\begin{align}
&  \frac{\omega^{2}}{4\Delta^{2}}\left(  \sqrt{V_{+}}\mathbf{B}_{1m_{j}}%
+\sqrt{V_{-}}\mathbf{B}_{3m_{j}}\right) \nonumber\\
&  -\sum_{lm_{j}^{\prime}}\left(  \sqrt{V_{+}}\left\langle \left(
\mathbf{\bar{b}b}_{1m_{j}}^{\ast}\right)  \left(  \mathbf{\bar{b}b}%
_{lm_{j}^{\prime}}\right)  \right\rangle +\sqrt{V_{-}}\left\langle \left(
\mathbf{\bar{b}b}_{3m_{j}}^{\ast}\right)  \left(  \mathbf{\bar{b}b}%
_{lm_{j}^{\prime}}\right)  \right\rangle \right)  \mathbf{B}_{lm_{j}^{\prime}%
}\nonumber\\
&  =-\frac{\omega}{2\Delta}i\left(  \sqrt{V_{+}}\left\langle \mathbf{b}%
_{1m_{j}}^{\ast}\times\mathbf{\bar{b}}\right\rangle +\sqrt{V_{-}}\left\langle
\mathbf{b}_{3m_{j}}^{\ast}\times\mathbf{\bar{b}}\right\rangle \right)  ,
\label{Bj2eq1}%
\end{align}%
\begin{align}
&  \frac{\omega^{2}}{4\Delta^{2}}\left(  \sqrt{V_{-}}\mathbf{B}_{1m_{j}}%
-\sqrt{V_{+}}\mathbf{B}_{3m_{j}}\right) \nonumber\\
&  -\sum_{lm_{j}^{\prime}}\left(  \sqrt{V_{-}}\left\langle \left(
\mathbf{\bar{b}b}_{1m_{j}}^{\ast}\right)  \left(  \mathbf{\bar{b}b}%
_{lm_{j}^{\prime}}\right)  \right\rangle -\sqrt{V_{+}}\left\langle \left(
\mathbf{\bar{b}b}_{3m_{j}}^{\ast}\right)  \left(  \mathbf{\bar{b}b}%
_{lm_{j}^{\prime}}\right)  \right\rangle \right)  \mathbf{B}_{lm_{j}^{\prime}%
}\nonumber\\
&  =-\frac{\omega}{2\Delta}i\left(  \sqrt{V_{-}}\left\langle \mathbf{b}%
_{1m_{j}}^{\ast}\times\mathbf{\bar{b}}\right\rangle -\sqrt{V_{+}}\left\langle
\mathbf{b}_{3m_{j}}^{\ast}\times\mathbf{\bar{b}}\right\rangle \right)  .
\label{Bj2eq2}%
\end{align}
Equations (\ref{Bj2eq1}), and (\ref{Bj2eq2}) represent a closed set of
equations for determining the vertex amplitudes $\mathbf{B}_{lm_{j}}\left(
\omega\right)  $.

\section{Eigenmodes of spin oscillations}

For further progress we have to define the ground state of the condensate
which is specified by the vector $\mathbf{\bar{b}}\left(  \mathbf{n}\right)
$. We focus on the condensation with $m_{j}=0$ which is conventionally
considered as the preferable one in the bulk matter of neutron stars
\cite{Elg}, \cite{Page04}, \cite{Page09}. In this case only the partial gap
amplitudes with $l=1,3$ and $m_{j}=0$ contribute. To simplify the notation we
denote them as $\Delta_{10}\equiv\Delta_{1}$ and $\Delta_{30}\equiv\Delta_{3}%
$. Taking into account spin-orbit and tensor interactions the ground state of
such a triplet condensate is given by the vector $\mathbf{\bar{b}}\left(
\mathbf{n}\right)  $ of the form
\begin{equation}
\mathbf{\bar{b}}\left(  \mathbf{n}\right)  =\frac{\Delta_{1}}{\Delta
}\mathbf{b}_{10}\left(  \mathbf{n}\right)  +\frac{\Delta_{3}}{\Delta
}\mathbf{b}_{30}\left(  \mathbf{n}\right)  , \label{bbar}%
\end{equation}
where $\Delta^{2}=\Delta_{1}^{2}+\Delta_{3}^{2}$. For this particular form of
the ground state the following relations can be verified by a straightforward
calculation%
\begin{equation}
\left\langle \left(  \mathbf{\bar{b}b}_{lm_{j}}^{\ast}\right)  \left(
\mathbf{\bar{b}b}_{l^{\prime}m_{j}^{\prime}}\right)  \right\rangle
=\delta_{m_{j}m_{j}^{\prime}}\left(  \mathbf{\bar{b}b}_{lm_{j}}^{\ast}\right)
\left(  \mathbf{\bar{b}b}_{l^{\prime}m_{j}}\right)  .~ \label{bbbbdfi}%
\end{equation}
We denote%
\begin{equation}
\beta_{m_{j}}^{l,l^{\prime}}\equiv\left\langle \left(  \mathbf{\bar{b}%
b}_{lm_{j}}^{\ast}\right)  \left(  \mathbf{\bar{b}b}_{l^{\prime}m_{j}}\right)
\right\rangle , \label{beta}%
\end{equation}
and%
\[
\Omega=\frac{\omega}{2\Delta}.
\]

From Eqs. (\ref{Bj2eq1}) and (\ref{Bj2eq2}) one can obtain five sets of linear
equations corresponding to different values of $m_{j}=0,\pm1,\pm2$.

For $m_{j}=0$ we obtain a set of two equations:%
\begin{gather}
\left[  \sqrt{V_{+}}\left(  \Omega^{2}-\beta_{0}^{1,1}\right)  -\sqrt{V_{-}%
}\beta_{0}^{3,1}\right]  \mathbf{B}_{10}\nonumber\\
+\left[  \sqrt{V_{-}}\left(  \Omega^{2}-\beta_{0}^{3,3}\right)  -\sqrt{V_{+}%
}\beta_{0}^{1,3}\right]  \mathbf{B}_{30}=0, \label{2mj0}%
\end{gather}%
\begin{gather}
\left[  \sqrt{V_{-}}\left(  \Omega^{2}-\beta_{0}^{1,1}\right)  +\sqrt{V_{+}%
}\beta_{0}^{3,1}\right]  \mathbf{B}_{10}\nonumber\\
-\left[  \sqrt{V_{+}}\left(  \Omega^{2}\mathbf{-}\beta_{0}^{3,3}\right)
+\sqrt{V_{-}}\beta_{0}^{1,3}\right]  \mathbf{B}_{30}=0, \label{3mj0}%
\end{gather}
The sets of equations for $m_{j}=\pm2$ are of the form%
\begin{gather}
\left[  \sqrt{V_{+}}\left(  \Omega^{2}-\beta_{\pm2}^{1,1}\right)  -\sqrt
{V_{-}}\beta_{\pm2}^{3,1}\right]  \mathbf{B}_{1,\pm2}\nonumber\\
+\left[  \sqrt{V_{-}}\left(  \Omega^{2}-\beta_{\pm2}^{3,3}\right)
-\sqrt{V_{+}}\beta_{\pm2}^{1,3}\right]  \mathbf{B}_{3,\pm2}=0, \label{1mj2}%
\end{gather}%
\begin{gather}
\left[  \sqrt{V_{-}}\left(  \Omega^{2}-\beta_{\pm2}^{1,1}\right)  +\sqrt
{V_{+}}\beta_{\pm2}^{3,1}\right]  \mathbf{B}_{1,\pm2}\nonumber\\
-\left[  \sqrt{V_{+}}\left(  \Omega^{2}-\beta_{\pm2}^{3,3}\right)
\mathbf{+}\sqrt{V_{-}}\beta_{\pm2}^{1,3}\right]  \mathbf{B}_{3,\pm2}=0.
\label{2mj2}%
\end{gather}
For each of the values of $m_{j}=\pm1$ we get a set of two equations:
\begin{align}
&  \left[  \sqrt{V_{+}}\left(  \Omega^{2}-\beta_{\pm1}^{1,1}\right)
-\sqrt{V_{-}}\beta_{\pm1}^{3,1}\right]  \mathbf{B}_{1,\pm1}\nonumber\\
&  +\left[  \sqrt{V_{-}}\left(  \Omega^{2}-\beta_{\pm1}^{3,3}\right)
-\sqrt{V_{+}}\beta_{\pm1}^{1,3}\right]  \mathbf{B}_{3,\pm1}\nonumber\\
&  =-\Omega i\left(  \sqrt{V_{+}}\left\langle \mathbf{b}_{1,\pm1}^{\ast}%
\times\mathbf{\bar{b}}\right\rangle +\sqrt{V_{-}}\left\langle \mathbf{b}%
_{3,\pm1}^{\ast}\times\mathbf{\bar{b}}\right\rangle \right)  , \label{2mj1}%
\end{align}%
\begin{align}
&  \left[  \sqrt{V_{-}}\left(  \Omega^{2}-\sqrt{V_{-}}\beta_{\pm1}%
^{1,1}\right)  +\sqrt{V_{+}}\beta_{\pm1}^{3,1}\right]  \mathbf{B}_{1,\pm
1}\nonumber\\
&  -\left[  \sqrt{V_{+}}\left(  \Omega^{2}-\beta_{\pm1}^{3,3}\right)
\mathbf{+}\sqrt{V_{-}}\beta_{\pm1}^{1,3}\right]  \mathbf{B}_{3,\pm
1}\nonumber\\
&  =-\Omega i\left(  \sqrt{V_{-}}\left\langle \mathbf{b}_{1,\pm1}^{\ast}%
\times\mathbf{\bar{b}}\right\rangle -\sqrt{V_{+}}\left\langle \mathbf{b}%
_{3,\pm1}^{\ast}\times\mathbf{\bar{b}}\right\rangle \right)  . \label{3mj1}%
\end{align}
By the same reason as in the case of Eqs. (\ref{Bm}) the uniform equations for
$m_{j}=0$ and for $m_{j}=\pm2$ have only trivial physical solutions
$\mathbf{B}_{10}=\mathbf{B}_{30}=0$ and $\mathbf{B}_{1,\pm2}=\mathbf{B}%
_{3,\pm2}=0$. The solutions to Eqs. (\ref{2mj1}) and (\ref{3mj1}) are
\begin{equation}
\mathbf{B}_{1,\pm1}=\frac{-i\Omega}{\chi_{\pm1}\left(  \Omega\right)  }\left[
\left(  \Omega^{2}-\beta_{\pm1}^{3,3}\right)  \left\langle \mathbf{b}_{1,\pm
1}^{\ast}\times\mathbf{\bar{b}}\right\rangle +\beta_{\pm1}^{1,3}\left\langle
\mathbf{b}_{3,\pm1}^{\ast}\times\mathbf{\bar{b}}\right\rangle \right]  ,
\label{B21}%
\end{equation}%
\begin{equation}
\mathbf{B}_{3,\pm1}=\frac{-i\Omega}{\chi_{\pm1}\left(  \Omega\right)  }\left[
\left(  \Omega^{2}-\beta_{\pm1}^{1,1}\right)  \left\langle \mathbf{b}_{3,\pm
1}^{\ast}\times\mathbf{\bar{b}}\right\rangle +\beta_{\pm1}^{3,1}\left\langle
\mathbf{b}_{1,\pm1}^{\ast}\times\mathbf{\bar{b}}\right\rangle \right]  ,
\label{B23}%
\end{equation}
where%
\begin{equation}
\left\langle \mathbf{b}_{1,\pm1}^{\ast}\mathbf{\times\bar{b}}\right\rangle
=\frac{1}{2}\sqrt{\frac{3}{2}}\frac{\Delta_{1}}{\Delta}\left(  i,\pm
1,0\right)  , \label{bcr1}%
\end{equation}%
\begin{equation}
\left\langle \mathbf{b}_{3,\pm1}^{\ast}\mathbf{\times\bar{b}}\right\rangle
=-\frac{1}{6}\sqrt{6}\frac{\Delta_{3}}{\Delta}\left(  i,\pm1,0\right)  ,
\label{bcr3}%
\end{equation}
and%
\begin{equation}
\chi_{\pm1}\left(  \Omega\right)  =\left(  \Omega^{2}-\beta_{\pm1}%
^{1,1}\right)  \left(  \Omega^{2}-\beta_{\pm1}^{3,3}\right)  -\beta_{\pm
1}^{1,3}\beta_{\pm1}^{3,1}. \label{hi}%
\end{equation}
Notice that the interaction parameters $V_{\pm}$ drop out of the final result
in Eqs. (\ref{B21}) and (\ref{B23}).

Poles of the effective vertices at $\chi_{\pm1}\left(  \Omega\right)  =0$
signal the existence of collective modes. Frequencies of the eigenoscillations
are independent of the sign of $m_{j}$. For each of the values of $m_{j}=\pm1$
we obtain two eigenmodes:
\begin{align}
\Omega_{1,2}^{2}  &  =\frac{1}{2}\left(  \beta_{m_{j}}^{1,1}+\beta_{m_{j}%
}^{3,3}\right. \nonumber\\
&  \left.  \pm\sqrt{\left(  \beta_{m_{j}}^{3,3}-\beta_{m_{j}}^{1,1}\right)
^{2}+4\beta_{m_{j}}^{1,3}\beta_{m_{j}}^{3,1}}\right)  . \label{w12}%
\end{align}
Straightforward calculations give%
\begin{equation}
\beta_{\pm1}^{1,1}=\frac{1}{20}\frac{\Delta_{1}^{2}}{\Delta^{2}}\left(
1-\frac{2}{7}\sqrt{6}\frac{\Delta_{3}}{\Delta_{1}}+\frac{58\Delta_{3}^{2}%
}{7\Delta_{1}^{2}}\right)  , \label{be11}%
\end{equation}%
\begin{equation}
\beta_{\pm1}^{3,3}=\frac{29}{70}\frac{\Delta_{1}^{2}}{\Delta^{2}}\left(
1-\frac{32}{87}\sqrt{6}\frac{\Delta_{3}}{\Delta_{1}}+\frac{28}{87}\frac
{\Delta_{3}^{2}}{\Delta_{1}^{2}}\right)  , \label{be33}%
\end{equation}%
\begin{equation}
\beta_{\pm1}^{1,3}=\beta_{\pm1}^{3,1}=-\frac{1}{140}\sqrt{6}\frac{\Delta
_{1}^{2}}{\Delta^{2}}\left(  1-11\sqrt{6}\frac{\Delta_{3}}{\Delta_{1}}%
+\frac{32}{3}\frac{\Delta_{3}^{2}}{\Delta_{1}^{2}}\right)  . \label{be13}%
\end{equation}

It is interesting to compare the resonant frequencies (\ref{w12}) with the
spin oscillation frequency $\omega_{s}=\Delta/\sqrt{5}~$obtained in Refs.
\cite{L09a,L10a} in a simple model restricted to excitations of the condensate
with $l=1$. By taking $\Delta_{3}=0$ and $\Delta_{1}=\Delta$ in the matrix
elements we find $4\beta_{\pm1}^{1,3}\beta_{\pm1}^{3,1}\ll\left(  \beta_{\pm
1}^{3,3}-\beta_{\pm1}^{1,1}\right)  ^{2}$. By neglecting the small term
$4\beta_{\pm1}^{1,3}\beta_{\pm1}^{3,1}$ under the root in Eq. (\ref{w12}) we
obtain%
\begin{equation}
\omega_{1}\left(  \Delta_{3}=0\right)  =2\Delta\beta_{\pm1}^{1,1}\left(
\Delta_{3}=0\right)  =\frac{1}{\sqrt{5}}\Delta\simeq0.45\Delta, \label{w1d0}%
\end{equation}
and%
\begin{equation}
\omega_{2}\left(  \Delta_{3}=0\right)  =2\Delta\beta_{\pm1}^{3,3}\left(
\Delta_{3}=0\right)  =\sqrt{\allowbreak\frac{58}{35}}\Delta~\simeq
1.\,\allowbreak29\Delta. \label{w2d0}%
\end{equation}
We see that the extending of the decomposition scheme of the excited states
with the total angular momentum $j=2$ up to $l=1,3$ leads to a very small
frequency shift of the known mode, $\omega_{s}=\Delta/\sqrt{5}$, but opens the
new additional mode of collective spin oscillations. Inclusion of the tensor
interaction implies $\Delta_{3}\neq0$ and $\Delta^{2}=\Delta_{1}^{2}%
+\Delta_{3}^{2}$. In this case from Eq. (\ref{w12}) we obtain two twofold
($m_{j}=\pm1$) frequency:%
\begin{align}
\omega_{1}^{2}  &  =\Delta_{1}^{2}\left(  \frac{13}{14}-\frac{1}{3}\sqrt
{6}\frac{\Delta_{3}}{\Delta_{1}}+\frac{23}{21}\frac{\Delta_{3}^{2}}{\Delta
_{1}^{2}}\right. \nonumber\\
&  \left.  -\sqrt{\frac{15}{28}-\frac{25}{49}\sqrt{6}\frac{\Delta_{3}}%
{\Delta_{1}}+\frac{485}{147}\frac{\Delta_{3}^{2}}{\Delta_{1}^{2}}%
-\allowbreak\frac{370}{441}\sqrt{6}\frac{\Delta_{3}^{3}}{\Delta_{1}^{3}}%
+\frac{55}{63}\frac{\Delta_{3}^{4}}{\Delta_{1}^{4}}}\right)  , \label{w3}%
\end{align}%
\begin{align}
\omega_{2}^{2}  &  =\Delta_{1}^{2}\left(  \frac{13}{14}-\frac{1}{3}\sqrt
{6}\frac{\Delta_{3}}{\Delta_{1}}+\frac{23}{21}\frac{\Delta_{3}^{2}}{\Delta
_{1}^{2}}\right. \nonumber\\
&  \left.  +\sqrt{\frac{15}{28}-\frac{25}{49}\sqrt{6}\frac{\Delta_{3}}%
{\Delta_{1}}+\frac{485}{147}\frac{\Delta_{3}^{2}}{\Delta_{1}^{2}}%
-\allowbreak\frac{370}{441}\sqrt{6}\frac{\Delta_{3}^{3}}{\Delta_{1}^{3}}%
+\frac{55}{63}\frac{\Delta_{3}^{4}}{\Delta_{1}^{4}}}\right)  . \label{w4}%
\end{align}
According to calculations of different authors, at the Fermi surface one has
$\Delta_{3}\simeq0.17\Delta_{1}$ (see, \textit{e.g.}, Ref. \cite{Khod}). In
this case our theoretical analysis predicts two degenerate modes with
$\omega=\omega_{1}\simeq0.42\Delta$ and two degenerate modes with
$\omega=\omega_{2}=1.\,\allowbreak19\Delta$.

\section{Anomalous vertices and polarization functions}

Making use of Eqs. (\ref{T1A}), (\ref{T2A}), (\ref{Bpm}), and (\ref{BpBm}) we
find%
\begin{equation}
\mathbf{\hat{T}}^{\left(  1\right)  }\left(  \mathbf{n},\omega\right)
=\sum_{m_{j}=\pm1}\left[  \mathbf{B}_{1m_{j}}\left(
\bm{\hat{\sigma}}\mathbf{b}_{1m_{j}}\right)  \hat{g}+\mathbf{B}_{3m_{j}%
}\left(  \bm{\hat{\sigma}}\mathbf{b}_{3m_{j}}\right)  \hat{g}\right]  ,
\label{T1B}%
\end{equation}%
\begin{equation}
\mathbf{\hat{T}}^{\left(  2\right)  }\left(  \mathbf{n},\omega\right)
=\sum_{m_{j}=\pm1}\left[  \mathbf{B}_{1m_{j}}\hat{g}\left(
\bm{\hat{\sigma}}\mathbf{b}_{1m_{j}}\right)  +\mathbf{B}_{3m_{j}}\hat
{g}\left(  \bm{\hat{\sigma}}\mathbf{b}_{3m_{j}}\right)  \right]  , \label{T2B}%
\end{equation}
where the functions $\mathbf{B}_{1m_{j}}\left(  \omega\right)  $
and$~\mathbf{B}_{3m_{j}}\left(  \omega\right)  $ are given in Eqs. (\ref{B21})
and (\ref{B23}).

The general expression of the axial polarization tensor in the BCS
approximation has been already discussed before. It can be obtained in the
form \cite{L09a,L10a,L10b}
\begin{align}
\Pi_{\mathrm{A}}^{ij}\left(  \omega\right)   &  =-4\varrho\mathcal{I}%
_{\mathsf{av}}\left(  \delta_{ij}-\left\langle \bar{b}_{i}\bar{b}%
_{j}\right\rangle \right) \nonumber\\
&  -\varrho\Omega\mathcal{I}_{\mathsf{av}}\left\langle \mathrm{Tr}\left[
\hat{\sigma}_{i}\hat{T}_{j}^{\left(  1\right)  }\hat{g}\left(  \hat
{\bm{\sigma}}\mathbf{\bar{b}}\right)  \right]  -~\mathrm{Tr}\left[
\hat{\sigma}_{i}\left(  \hat{\bm{\sigma}}\mathbf{\bar{b}}\right)  \hat{g}%
\hat{T}_{j}^{\left(  2\right)  }\right]  \right\rangle , \label{KA}%
\end{align}
where $i,j=1,2,3$ and the anomalous axial-vector vertices $\mathbf{\hat{T}%
}^{\left(  1,2\right)  }$ are given by Eqs. (\ref{T1B}) and (\ref{T2B}). We
omit for brevity the dependence on $\mathbf{n}$ and $\omega$. Calculation of
the traces results in the expression%
\begin{align}
\Pi_{\mathrm{A}}^{ij}\left(  \omega\right)   &  =-4\varrho\mathcal{I}%
_{\mathsf{av}}\left(  \delta^{ij}-\left\langle \bar{b}^{i}\bar{b}%
^{j}\right\rangle \right) \nonumber\\
&  +4\varrho\Omega\mathcal{I}_{\mathsf{av}}\sum_{m_{j}=\pm1}\left(
i\left\langle \mathbf{b}_{1m_{j}}\mathbf{\times\bar{b}}\right\rangle
^{i}B_{1m_{j}}^{j}+i\left\langle \mathbf{b}_{3m_{j}}\mathbf{\times\bar{b}%
}\right\rangle ^{i}B_{3m_{j}}^{j}\right)  , \label{PiA}%
\end{align}
Inserting the vectors $\mathbf{B}_{l,m_{j}}$, given by Eqs. (\ref{B21}), and
(\ref{B23}), we write the function (\ref{hi}), as%
\begin{equation}
\chi_{m_{j}}\left(  \Omega\right)  =\left(  \Omega^{2}-\Omega_{1}^{2}\right)
\left(  \Omega^{2}-\Omega_{2}^{2}\right)  , \label{hiw}%
\end{equation}
where $m_{j}=\pm1$, and%
\begin{equation}
\Omega_{1}^{2}=\frac{\omega_{1}^{2}}{4\Delta^{2}}\,,~\Omega_{2}^{2}%
=\frac{\omega_{2}^{2}}{4\Delta^{2}}.~ \label{W34}%
\end{equation}
In this way we obtain%
\begin{align}
\Pi_{\mathrm{A}}^{ij}\left(  \omega\right)   &  =-4\varrho\mathcal{I}%
_{\mathsf{av}}\left(  \delta^{ij}-\left\langle \bar{b}^{i}\bar{b}%
^{j}\right\rangle \right) \nonumber\\
&  +4\varrho\mathcal{I}_{\mathsf{av}}\frac{\Omega^{2}}{\left(  \Omega
^{2}-\Omega_{1}^{2}+i0\right)  \left(  \Omega^{2}-\Omega_{2}^{2}+i0\right)
}\nonumber\\
&  \times\sum_{m_{j}=\pm1}\left[  \left(  \Omega^{2}-\beta_{m_{j}}%
^{3,3}\right)  \left\langle \mathbf{b}_{1m_{j}}\mathbf{\times\bar{b}%
}\right\rangle ^{i}\left\langle \mathbf{b}_{1m_{j}}^{\ast}\times
\mathbf{\bar{b}}\right\rangle ^{j}\right. \nonumber\\
&  +\beta_{m_{j}}^{1,3}\left\langle \mathbf{b}_{1m_{j}}\mathbf{\times\bar{b}%
}\right\rangle ^{i}\left\langle \mathbf{b}_{3m_{j}}^{\ast}\times
\mathbf{\bar{b}}\right\rangle ^{j}\nonumber\\
&  +\left(  \Omega^{2}-\beta_{m_{j}}^{1,1}\right)  \left\langle \mathbf{b}%
_{3m_{j}}\mathbf{\times\bar{b}}\right\rangle ^{i}\left\langle \mathbf{b}%
_{3m_{j}}^{\ast}\times\mathbf{\bar{b}}\right\rangle ^{j}\nonumber\\
&  \left.  +\beta_{m_{j}}^{3,1}\left\langle \mathbf{b}_{3m_{j}}\mathbf{\times
\bar{b}}\right\rangle ^{i}\left\langle \mathbf{b}_{1m_{j}}^{\ast}%
\times\mathbf{\bar{b}}\right\rangle ^{j}\right]  . \label{Pa}%
\end{align}
The poles location on the complex plane of $\Omega$ is chosen so that to
obtain the retarded polarization function.

Summation over $m_{j}=\pm1$ can be done using the fact that the parameters
$\beta_{m_{j}}^{l,l^{\prime}}$ entering this equation are independent of the
sign of $m_{j}$, and $\beta_{\pm1}^{1,3}=\beta_{\pm1}^{3,1}$. Then a simple
calculation gives%
\begin{equation}
\sum_{m_{j}=\pm1}\left\langle \mathbf{b}_{lm_{j}}\mathbf{\times\bar{b}%
}\right\rangle ^{i}\left\langle \mathbf{b}_{l^{\prime}m_{j}}^{\ast}%
\times\mathbf{\bar{b}}\right\rangle ^{j}=\lambda_{ll^{\prime}}\left(
\delta_{ij}-\delta_{i3}\delta_{j3}\right)  \label{lamll}%
\end{equation}
with%
\begin{equation}
\lambda_{11}=\frac{3}{4}\frac{\Delta_{1}^{2}}{\Delta^{2}}\mathsf{~}%
,\mathsf{~}\lambda_{33}=\frac{1}{3}\frac{\Delta_{3}^{2}}{\Delta^{2}%
}\,,~\lambda_{13}=\lambda_{31}=-\frac{1}{2}\frac{\Delta_{1}\Delta_{3}}%
{\Delta^{2}}, \label{Lam11}%
\end{equation}
and%
\begin{equation}
\left\langle \bar{b}^{i}\bar{b}^{j}\right\rangle =\frac{1}{6}\frac{\Delta
_{1}^{2}}{\Delta^{2}}\left[  \left(  1+\frac{12}{7}\frac{\Delta_{3}^{2}%
}{\Delta_{1}^{2}}\right)  \delta^{ij}+\left(  3+\frac{6}{7}\frac{\Delta
_{3}^{2}}{\Delta_{1}^{2}}\right)  \delta^{i3}\delta^{3j}\right]  . \label{bb}%
\end{equation}
We finally obtain the expression%
\begin{align}
\Pi_{\mathrm{A}}^{ij}\left(  \omega\right)   &  =-4\varrho\mathcal{I}%
_{\mathsf{av}}\left(  \delta^{ij}-\left\langle \bar{b}^{i}\bar{b}%
^{j}\right\rangle \right) \nonumber\\
&  +4\varrho\mathcal{I}_{\mathsf{av}}\left(  \delta^{ij}-\delta^{i3}%
\delta^{3j}\right)  \frac{4\Delta^{2}\omega^{2}}{\left(  \omega^{2}-\omega
_{1}^{2}+i0\right)  \left(  \omega^{2}-\omega_{2}^{2}+i0\right)  }\nonumber\\
&  \times\left[  \lambda_{11}\left(  \frac{\omega^{2}}{4\Delta^{2}}-\beta
_{1}^{3,3}\right)  +\lambda_{33}\left(  \frac{\omega^{2}}{4\Delta^{2}}%
-\beta_{1}^{1,1}\right)  +2\lambda_{13}\beta_{1}^{1,3}\right]  . \label{PA}%
\end{align}

Below we use the retarded polarization tensor for\ a calculation of the
neutrino emissivity of \ a nonrelativistic superfluid matter. In this
calculation one can neglect the temporal and mixed components of the
polarization tensor occurring as small relativistic corrections.

\section{Neutrino energy losses$\allowbreak\allowbreak$}

We examine the neutrino energy losses in the standard model of weak
interactions. Then after integration over the phase volume of freely escaping
neutrinos and antineutrinos the total energy which is emitted per unit volume
and time can be obtained in the form (see details, e.g., in Ref. \cite{L01})
\begin{equation}
\epsilon=-\frac{G_{F}^{2}C_{A}^{2}\mathcal{N}_{\nu}}{192\pi^{5}}\int
_{0}^{\infty}d\omega\int d^{3}q\frac{\omega\Theta\left(  \omega-q\right)
}{\exp\left(  \frac{\omega}{T}\right)  -1}\operatorname{Im}\Pi_{\mathrm{A}%
}^{\mu\nu}\left(  \omega,\mathbf{q}\right)  \left(  k_{\mu}k_{\nu}-k^{2}%
g_{\mu\nu}\right)  , \label{QQQ}%
\end{equation}
where $G_{F}$ is the Fermi coupling constant, $C_{A}=1.26$ is the axial-vector
weak coupling constant of neutrons, $\mathcal{N}_{\nu}=3$ is the number of
neutrino flavors, $\Theta\left(  x\right)  $ is the Heaviside step function,
and $k^{\mu}=\left(  \omega,\mathbf{q}\right)  $ is the total energy and
momentum of the freely escaping neutrino pair $\left(  \mu,\nu=0,1,2,3\right)
$.

In Eq. (\ref{QQQ}), we have neglected the neutrino emission in the vector
channel, which is strongly suppressed owing to conservation of the vector
current \cite{LP06}, \cite{L09a}. Therefore the energy losses are connected to
the imaginary part of the retarded polarization tensor in the axial channel,
$\operatorname{Im}\Pi_{\mathrm{A}}^{\mu\nu}\simeq\delta^{\mu i}\delta^{\nu
j}\operatorname{Im}\Pi_{\mathrm{A}}^{ij}$. The latter is caused by the PBF
processes and by the SWDs. These processes operate in different kinematical
domains, so that the imaginary part of the polarization tensor consists of two
clearly distinguishable contributions, $\operatorname{Im}\Pi_{\mathrm{A}}%
^{ij}=\operatorname{Im}\Pi_{\mathrm{PBF}}^{ij}+\operatorname{Im}%
\Pi_{\mathrm{SWD}}^{ij}$, which we now consider.

\subsection{PBF channel}

The imaginary part of $\mathcal{I}_{\mathsf{av}}$, which arises from the poles
of the integrand in Eq. (\ref{Iav}) at $\left\vert \omega\right\vert =2E$ is
given by%
\begin{equation}
\operatorname{Im}\mathcal{I}_{\mathsf{av}}\left(  \omega>0\right)  =\frac{\pi
}{2}\frac{\Delta^{2}\Theta\left(  \omega^{2}-4\Delta^{2}\right)  }{\omega
\sqrt{\omega^{2}-4\Delta^{2}}}\tanh\left(  \frac{\omega}{4T}\right)  .
\label{ImI}%
\end{equation}
With the aid of this expression we find%
\begin{subequations}
\begin{align}
\operatorname{Im}\Pi_{ij}^{\mathrm{PBF}}\left(  \omega\right)   &
=-2\pi\varrho\frac{\Delta^{2}\Theta\left(  \omega^{2}-4\Delta^{2}\right)
}{\omega\sqrt{\omega^{2}-4\Delta^{2}}}\tanh\left(  \frac{\omega}{4T}\right)
\nonumber\\
\times &  \left\{  \delta_{ij}-\left\langle \bar{b}_{i}\bar{b}_{j}%
\right\rangle -\left(  \delta_{ij}-\delta_{i3}\delta_{3j}\right)  \frac{3}%
{4}\frac{\Delta_{1}^{2}}{\Delta^{2}}\frac{4\Delta^{2}\omega^{2}}{\left(
\omega^{2}-\omega_{1}^{2}\right)  \left(  \omega^{2}-\omega_{2}^{2}\right)
}\right. \nonumber\\
&  \left.  \times\left[  \left(  \frac{\omega^{2}}{4\Delta^{2}}-\beta
_{1}^{3,3}\right)  +\frac{4}{9}\frac{\Delta_{3}^{2}}{\Delta_{1}^{2}}\left(
\frac{\omega^{2}}{4\Delta^{2}}-\beta_{1}^{1,1}\right)  -\frac{4}{3}%
\frac{\Delta_{3}}{\Delta_{1}}\beta_{1}^{1,3}\right]  \right\}  . \label{ImPBF}%
\end{align}

Inserting the imaginary part of the polarization tensor into Eq. (\ref{QQQ}),
we calculate the contraction of $\operatorname{Im}\Pi_{\mathrm{PBF}}^{\mu\nu}$
with the symmetric tensor $k_{\mu}k_{\nu}-k^{2}g_{\mu\nu}$. This gives%
\end{subequations}
\begin{align}
\epsilon_{\mathrm{PBF}}  &  =\frac{1}{96\pi^{6}}G_{F}^{2}C_{A}^{2}%
\mathcal{N}_{\nu}p_{F}M^{\ast}\Delta^{2}\int_{2\Delta}^{\infty}d\omega\frac
{1}{\sqrt{\omega^{2}-4\Delta^{2}}}\frac{1}{\exp\left(  \frac{\omega}%
{T}\right)  -1}\tanh\left(  \frac{\omega}{4T}\right) \nonumber\\
&  \times\int\limits_{q<\omega}d^{3}q\left\{  2\omega^{2}-q^{2}-\frac{1}%
{6}\frac{\Delta_{1}^{2}}{\Delta^{2}}\left[  \left(  1+\frac{12}{7}\frac
{\Delta_{3}^{2}}{\Delta_{1}^{2}}\right)  q_{\perp}^{2}+\left(  4+\frac{18}%
{7}\frac{\Delta_{3}^{2}}{\Delta_{1}^{2}}\right)  q_{z}^{2}\right]  \right.
\nonumber\\
&  -\frac{3}{4}\left(  2\left(  \omega^{2}-q_{z}^{2}\right)  -q_{\perp}%
^{2}\right)  \frac{\Delta_{1}^{2}}{\Delta^{2}}\frac{4\Delta^{2}\omega^{2}%
}{\left(  \omega^{2}-\omega_{1}^{2}\right)  \left(  \omega^{2}-\omega_{2}%
^{2}\right)  }\nonumber\\
&  \left.  \times\left[  \left(  \frac{\omega^{2}}{4\Delta^{2}}-\beta
_{1}^{3,3}\right)  +\frac{4}{9}\frac{\Delta_{3}^{2}}{\Delta_{1}^{2}}\left(
\frac{\omega^{2}}{4\Delta^{2}}-\beta_{1}^{1,1}\right)  -\frac{4}{3}%
\frac{\Delta_{3}}{\Delta_{1}}\beta_{1}^{1,3}\right]  \right\}  . \label{Enu}%
\end{align}
Integration over $d^{3}q$ can be done in cylindrical frame, where
$q_{1}=q_{\perp}\cos\Phi$, $q_{2}=q_{\perp}\sin\Phi$, and $q_{3}=q_{z}$. This
results in the neutrino energy losses of the form%
\begin{align}
\epsilon_{\mathrm{PBF}}  &  =\frac{2}{3\pi^{5}}G_{F}^{2}C_{A}^{2}%
\mathcal{N}_{\nu}p_{F}M^{\ast}T^{7}y^{2}\int_{0}^{\infty}\frac{z^{4}%
dx}{\left(  1+\exp z\right)  ^{2}}\nonumber\\
&  \times\left\{  \frac{4}{5}-\frac{3}{5}\frac{\Delta_{1}^{2}}{\Delta^{2}%
}\frac{x^{2}+y^{2}}{\left(  x^{2}+y^{2}\left(  1-\Omega_{1}^{2}\right)
\right)  \left(  x^{2}+y^{2}\left(  1-\Omega_{2}^{2}\right)  \right)  }\right.
\nonumber\\
&  \times\left.  \left[  x^{2}+y^{2}\left(  1-\beta_{1}^{3,3}\right)
+\frac{4}{9}\frac{\Delta_{3}^{2}}{\Delta_{1}^{2}}\left(  x^{2}+y^{2}\left(
1-\beta_{1}^{1,1}\right)  \right)  -\frac{4}{3}\frac{\Delta_{3}}{\Delta_{1}%
}y^{2}\beta_{1}^{1,3}\right]  \right\}  , \label{PBF}%
\end{align}
where $z=\sqrt{x^{2}+y^{2}}$, $y=\Delta\left(  T\right)  /T$, and
$\Omega_{1,2}$ are defined in Eq. (\ref{W34}). \allowbreak\allowbreak
\allowbreak In obtaining Eq. (\ref{PBF}) the change is used $\omega
=2T\sqrt{x^{2}+\Delta^{2}/T^{2}}$.

For a practical usage from Eq. (\ref{PBF}), we find%
\begin{equation}
\epsilon_{\mathrm{PBF}}=5.\,\allowbreak85\times10^{20}~\left(  \frac{M^{\ast}%
}{M}\right)  \left(  \frac{p_{F}}{Mc}\right)  T_{9}^{7}\mathcal{N}_{\nu
}C_{\mathrm{A}}^{2}\mathcal{F}_{\mathrm{PBF}}\left(  y\right)  ~~~\frac
{\mathsf{erg}}{\mathsf{cm}^{3}\mathsf{s}}, \label{ergPBF}%
\end{equation}
~ where $M$ is the bare nucleon mass; $T_{9}=T/10^{9}\mathrm{K}$, and
\begin{align}
&  \mathcal{F}_{\mathrm{PBF}}\left(  y\right)  =y^{2}\int_{0}^{\infty}%
dx\frac{z^{4}}{\left(  1+\exp z\right)  ^{2}}\nonumber\\
&  \times\left\{  4-3\frac{\Delta_{1}^{2}}{\Delta^{2}}\frac{x^{2}+y^{2}%
}{\left(  x^{2}+y^{2}\left(  1-\Omega_{1}^{2}\right)  \right)  \left(
x^{2}+y^{2}\left(  1-\Omega_{2}^{2}\right)  \right)  }\right. \nonumber\\
&  \left.  \times\left[  x^{2}+y^{2}\left(  1-\beta_{1}^{3,3}\right)
+\frac{4}{9}\frac{\Delta_{3}^{2}}{\Delta_{1}^{2}}\left(  x^{2}+y^{2}\left(
1-\beta_{1}^{1,1}\right)  \right)  -\frac{4}{3}\frac{\Delta_{3}}{\Delta_{1}%
}y^{2}\beta_{1}^{1,3}\right]  \right\}  . \label{Ft}%
\end{align}

In the limit $\Delta_{3}=0$, the neutrino energy losses, as given by Eq.
(\ref{PBF}) reproduce the result obtained in Ref. \cite{L10b} for the
one-component phase $m_{j}=0$. It is necessary to notice that Eq. (\ref{Ft})
obtained in the angle-average approximation is much simpler for numerical
evaluations than the "exact" expression which contains additionally the angle
integration \cite{L10a}. To avoid possible misunderstanding we stress that the
gap amplitude $\Delta\left(  T\right)  $ in Eq. (\ref{Ft}) is $\sqrt{2}$ times
larger than the gap amplitude $\Delta_{YKL}$ used in Ref. \cite{YKL} , where
the same anisotropic gap $\Delta_{\mathbf{n}}=\Delta\bar{b}\left(
\mathbf{n}\right)  $ is written in the form $\Delta_{\mathbf{n}}=\Delta
_{YKL}\sqrt{1+3\cos^{2}\theta}\equiv\Delta_{YKL}\sqrt{2}\,\bar{b}\left(
\mathbf{n}\right)  $. In other words, $\left\langle \Delta_{\mathbf{n}}%
^{2}\right\rangle =\Delta^{2}=2\Delta_{YKL}^{2}$.

\subsection{SWD channel}

In the frequency domain $0<\omega<2\Delta$, the imaginary part of the weak
polarization tensor (\ref{PA}) arises from the poles of the denominator at
$\omega=\omega_{1}$ and $\omega=\omega_{2}$ and consists of two terms%
\begin{align}
\operatorname{Im}\Pi_{ij}^{\mathrm{SWD}}\left(  \omega>0\right)   &
=\nonumber\\
&  -2\pi\frac{3}{4}\varrho\left(  \delta^{ij}-\delta^{i3}\delta^{3j}\right)
\mathcal{I}_{\mathsf{av}}\left(  \omega_{1}\right)  \frac{\Delta_{1}^{2}%
}{\Delta^{2}}\frac{\omega_{1}\delta\left(  \omega-\omega_{1}\right)  }{\left(
\Omega_{1}^{2}-\Omega_{2}^{2}\right)  }\nonumber\\
&  \times\left[  \Omega_{1}^{2}-\beta_{1}^{3,3}+\frac{4}{9}\frac{\Delta
_{3}^{2}}{\Delta_{1}^{2}}\left(  \Omega_{1}^{2}-\beta_{1}^{1,1}\right)
-\frac{4}{3}\frac{\Delta_{3}}{\Delta_{1}}\beta_{1}^{1,3}\right] \nonumber\\
&  -2\pi\frac{3}{4}\varrho\left(  \delta^{ij}-\delta^{i3}\delta^{3j}\right)
\mathcal{I}_{\mathsf{av}}\left(  \omega_{2}\right)  \frac{\Delta_{1}^{2}%
}{\Delta^{2}}\frac{\omega_{2}\delta\left(  \omega-\omega_{2}\right)  }{\left(
\Omega_{2}^{2}-\Omega_{1}^{2}\right)  }\nonumber\\
&  \times\left[  \Omega_{2}^{2}-\beta_{1}^{3,3}+\frac{4}{9}\frac{\Delta
_{3}^{2}}{\Delta_{1}^{2}}\left(  \Omega_{2}^{2}-\beta_{1}^{1,1}\right)
-\frac{4}{3}\frac{\Delta_{3}}{\Delta_{1}}\beta_{1}^{1,3}\right]  .
\label{ImSWD}%
\end{align}
According to Eqs. (\ref{w2d0}) at $\Delta_{3}=0$ one has $\Omega_{2}%
=\beta_{\pm1}^{3,3}$ and $\operatorname{Im}\Pi_{ij}^{\mathrm{SWD}}\left(
\omega\rightarrow\omega_{2}\right)  =0$. In other words, the high-frequency
spin oscillations can not be excited if the tensor interactions between the
pairing particles are not taken into account.

Inserting Eq. (\ref{ImSWD}) into Eq. (\ref{QQQ}) and performing trivial
calculations, we find two contributions to the neutrino energy losses. The
first contribution is caused by the decay into neutrino pairs of the lowest
mode of spin oscillations at $\omega=\omega_{1}$:
\begin{align}
\epsilon_{\mathrm{SWD}}^{\left(  1\right)  }  &  =\frac{1}{320\pi^{5}}%
G_{F}^{2}C_{A}^{2}\mathcal{N}_{\nu}p_{F}M^{\ast}\frac{\Delta_{1}^{2}}%
{\Delta^{2}}\frac{1}{\Omega_{1}^{2}-\Omega_{2}^{2}}\nonumber\\
&  \times\left(  \Omega_{1}^{2}-\beta_{1}^{3,3}-\frac{4}{3}\frac{\Delta_{3}%
}{\Delta_{1}}\beta_{1}^{1,3}+\frac{4}{9}\frac{\Delta_{3}^{2}}{\Delta_{1}^{2}%
}\left(  \Omega_{1}^{2}-\beta_{1}^{1,1}\right)  \right) \nonumber\\
&  \times\frac{\omega_{1}^{7}}{\exp\left(  \frac{\omega_{1}}{T}\right)
-1}\int_{0}^{\infty}\frac{d\varepsilon}{E}\ \frac{\Delta^{2}}{E^{2}-\omega
_{1}^{2}/4}\tanh\frac{E}{2T}~. \label{Eswd1}%
\end{align}
According to this equation, in the case of $\Delta_{3}\rightarrow0$,
$\Delta_{1}\rightarrow\Delta$, and $\omega_{1}\rightarrow\Delta/\sqrt{5}$, the
energy losses are twice less than that found in Ref. (\cite{L10b}):
\begin{align}
\epsilon_{\mathrm{SWD}}^{\left(  1\right)  }\left(  \Delta_{3}=0\right)   &
=\frac{1}{320\pi^{5}}G_{F}^{2}C_{A}^{2}\mathcal{N}_{\nu}p_{F}M^{\ast
}\nonumber\\
&  \times\frac{\omega_{1}^{7}}{\exp\left(  \frac{\omega_{1}}{T}\right)
-1}\int_{0}^{\infty}\frac{d\varepsilon}{E}\ \frac{\Delta^{2}}{E^{2}-\omega
_{1}^{2}/4}\tanh\frac{E}{2T}. \label{SWD0}%
\end{align}
\emph{We use the opportunity to point out the error in Eqs. (79) and (81) of
Ref.} (\cite{L10b})\emph{, where the factor of }$1/2$\emph{ is lost. }

The second contribution originates from weak decays of the second (higher)
mode at $\omega=\omega_{2}$:
\begin{align}
\epsilon_{\mathrm{SWD}}^{\left(  2\right)  }  &  =\frac{1}{320\pi^{5}}%
G_{F}^{2}C_{A}^{2}\mathcal{N}_{\nu}p_{F}M^{\ast}\frac{\Delta_{1}^{2}}%
{\Delta^{2}}\frac{1}{\Omega_{2}^{2}-\Omega_{1}^{2}}\nonumber\\
&  \times\left(  \Omega_{2}^{2}-\beta_{1}^{3,3}-\frac{4}{3}\frac{\Delta_{3}%
}{\Delta_{1}}\beta_{1}^{1,3}+\frac{4}{9}\frac{\Delta_{3}^{2}}{\Delta_{1}^{2}%
}\left(  \Omega_{2}^{2}-\beta_{1}^{1,1}\right)  \right) \nonumber\\
\times &  \frac{\omega_{2}^{7}}{\exp\left(  \frac{\omega_{2}}{T}\right)
-1}\int_{0}^{\infty}\frac{d\varepsilon}{E}\ \frac{\Delta^{2}}{E^{2}-\omega
_{2}^{2}/4}\tanh\frac{E}{2T}. \label{Eswd2}%
\end{align}

Because excitation of the high-frequency spin oscillations occurs through the
tensor interactions, the contribution of the second mode vanishes if the
tensor forces are switched off (i.e. when $\Delta_{3}=0$) .

The expressions (\ref{Eswd1}) and (\ref{Eswd2}) can be written in the
traditional form
\begin{align}
\epsilon_{\mathrm{SWD}}^{\left(  1\right)  }  &  =1.\,\allowbreak
76\times10^{21}~\left(  \frac{M^{\ast}}{M}\right)  \left(  \frac{p_{F}}%
{Mc}\right)  \mathcal{N}_{\nu}C_{\mathrm{A}}^{2}T_{9}^{7}y^{7}\frac{\Delta
_{1}^{2}}{\Delta^{2}}\nonumber\\
&  \times\left(  \Omega_{1}^{2}-\beta_{1}^{3,3}-\frac{4}{3}\frac{\Delta_{3}%
}{\Delta_{1}}\beta_{1}^{1,3}+\frac{4}{9}\frac{\Delta_{3}^{2}}{\Delta_{1}^{2}%
}\left(  \Omega_{1}^{2}-\beta_{1}^{1,1}\right)  \right) \nonumber\\
&  \times\frac{1}{\Omega_{1}^{2}-\Omega_{2}^{2}}\ \frac{\Omega_{1}^{7}}%
{\exp\left(  2y\Omega_{1}\right)  -1}\mathcal{F}_{\mathrm{SWD}}\left(
\Omega_{1},y\right)  ~~\ \ \ \ \ \ \ \ ~~\frac{\mathsf{erg}}{\mathsf{cm}%
^{3}\mathsf{s}}, \label{ergSWD1}%
\end{align}%
\begin{align}
\epsilon_{\mathrm{SWD}}^{\left(  2\right)  }  &  =1.\,\allowbreak
76\times10^{21}~\left(  \frac{M^{\ast}}{M}\right)  \left(  \frac{p_{F}}%
{Mc}\right)  \mathcal{N}_{\nu}C_{\mathrm{A}}^{2}T_{9}^{7}y^{7}\frac{\Delta
_{1}^{2}}{\Delta^{2}}\nonumber\\
&  +\left(  \Omega_{2}^{2}-\beta_{1}^{3,3}-\frac{4}{3}\frac{\Delta_{3}}%
{\Delta_{1}}\beta_{1}^{1,3}+\frac{4}{9}\frac{\Delta_{3}^{2}}{\Delta_{1}^{2}%
}\left(  \Omega_{2}^{2}-\beta_{1}^{1,1}\right)  \right) \nonumber\\
&  \times\frac{1}{\Omega_{2}^{2}-\Omega_{1}^{2}}\ \frac{\Omega_{2}^{7}}%
{\exp\left(  2y\Omega_{2}\right)  -1}\mathcal{F}_{\mathrm{SWD}}\left(
\Omega_{2},y\right)  ~~\ \ \ \ \ \ \ \ ~~\frac{\mathsf{erg}}{\mathsf{cm}%
^{3}\mathsf{s}}, \label{ergSWD2}%
\end{align}
where $\beta_{1}^{l,l^{\prime}}$, given by Eqs. (\ref{be11})-(\ref{be13}), are
functions of the gap components $\Delta_{1}$ and $\Delta_{3}$; $y\equiv
\Delta\left(  T\right)  /T$, and%
\begin{equation}
\mathcal{F}_{\mathrm{SWD}}\left(  \Omega,y\right)  =y^{7}\int_{0}^{\infty
}\frac{du}{\sqrt{u^{2}+1}}\ \frac{1}{u^{2}+1-\Omega^{2}}\tanh\frac{y}{2}%
\sqrt{u^{2}+1}. \label{Fswd}%
\end{equation}

\section{Efficiency of the\ neutrino emission}

In general, the temperature dependence of the gap amplitudes $\Delta_{1}$ and
$\Delta_{3}$ is to be found with the aid of the gap equations. For simple
estimates we take the approximation that the ratio $\Delta_{3}/\Delta_{1}$
remains constant when the temperature varies, and the temperature dependence
of the gap is given by the function $y=\Delta\left(  T\right)  /T$. This
function is well investigated for a $^{3}P_{2}$ pairing. Since the tensor
contribution can be considered as a perturbation \cite{Khodel}, in a zero
approximation, we can use, for example, the simple fit to $\Delta_{YKL}\left(
T\right)  /T=\mathsf{v}_{B}\left(  \tau\right)  $, as suggested in Ref.
\cite{YKL}, where $\tau\equiv T/T_{c}$. Taking into account that, in Ref.
\cite{YKL}, the gap amplitude $\Delta_{YKL}\left(  T\right)  $ is defined by
the relation $\Delta_{\mathbf{n}}^{2}=\Delta_{YKL}^{2}\left(  1+3\cos
^{2}\theta\right)  $, while our definition is $\Delta_{\mathbf{n}}^{2}\left(
\Delta_{3}=0\right)  =\frac{1}{2}\Delta^{2}\left(  1+3\cos^{2}\theta\right)
$, we obtain $y\left(  \tau\right)  =\sqrt{2}\mathsf{v}_{B}\left(
\tau\right)  $.

\begin{figure}[ptb]
\includegraphics{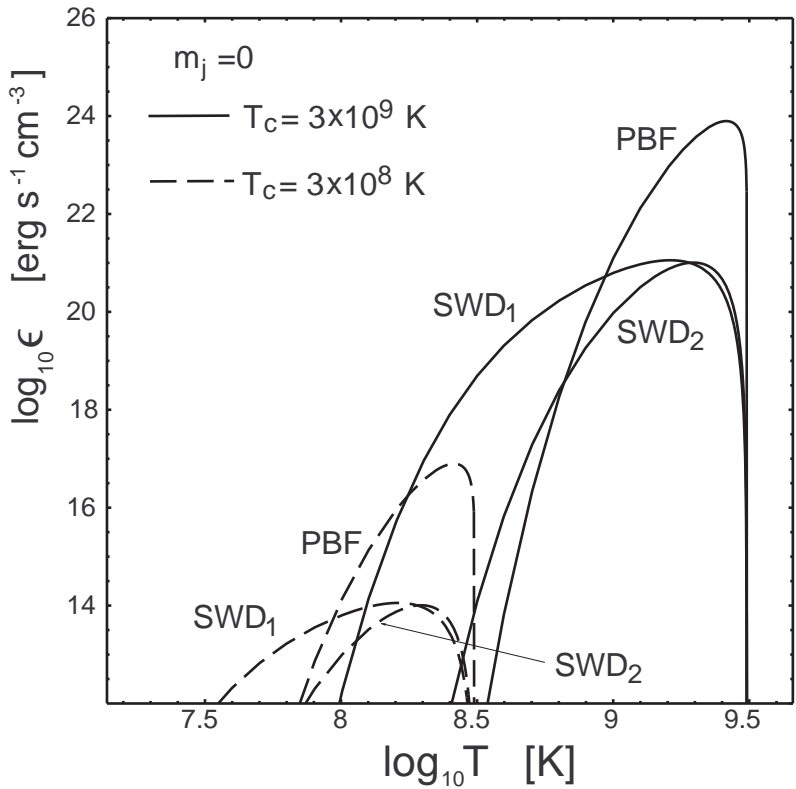}\caption{Temperature dependence of the neutrino
emissivity owing to recombination of Cooper pairs (PBF) and owing to decay of
spin waves (SWD$_{1}$, the emissivity of the lower mode. SWD$_{2}$, the
emissivity of the upper mode) at $\Delta_{3}/\Delta_{1}=0.17$ and
$p_{F}=2.1~fm^{-1}$. }%
\label{fig2}%
\end{figure}

In Fig. \ref{fig2} we show the neutrino emissivity $\epsilon$ caused by the
PBF processes and by the decay of the lowest mode (SWD$_{1}$) and the higher
mode (SWD$_{2}$) of spin oscillations. The temperature dependence of the
emissivity is evaluated at $p_{F}=2.1fm^{-1}$ assuming $\Delta_{3}/\Delta_{1}%
$\noindent$~=$\noindent$~0.17$. We set the effective nucleon masses $M^{\ast
}=0.7M$; the critical temperature for neutron pairing is chosen to be
$T_{c}=3\times10^{9}K$ and $T_{c}=3\times10^{8}K$.

One can see that the decay of the low-frequency spin waves into neutrino pairs
(SWD$_{1}$) is very effective at low temperatures, when other known mechanisms
of neutrino energy losses in the bulk neutron matter are strongly suppressed
by the superfluidity. As discussed in Ref. \cite{L10b} the neutrino emission
caused by the decay of the low-frequency spin-waves can dominate the $\gamma$
radiation within a wide range of low temperatures, which was considered before
as the photon-cooling era. A simple estimate has shown that the decays of spin
waves$~$can modify the cooling trajectory of neutron stars (see Fig. 5 in Ref.
\cite{L10b}).

Weak decays of the high-frequency mode of spin oscillations occurs only if the
tensor forces are taken into account in the pairing interaction, that is, if
$\Delta_{3}\neq0$. Although the maximal neutrino emission caused by the
SWD$_{2}$ processes is as large as in the SWD$_{1}$ the neutrino energy losses
from the decay of the upper mode decrease more rapidly along with lowering of
the temperature. As a result the SWD$_{2}$ contribution into the total energy
losses is negligible in comparison with the sum of the PBF and SWD$_{1}$
contributions. We found that the latter can be excellently described by the
expressions obtained in Ref. \cite{L10b} for the case of $^{3}P_{2}$ pairing
with $m_{j}=0$ [However see note after Eq. (\ref{SWD0})]. We quote these
simple expressions for references:
\begin{equation}
\epsilon_{\mathrm{PBF}}=5.\,\allowbreak85\times10^{20}~\left(  \frac{M^{\ast}%
}{M}\right)  \left(  \frac{p_{F}}{Mc}\right)  T_{9}^{7}\mathcal{N}_{\nu
}C_{\mathrm{A}}^{2}F_{\mathrm{PBF}}\left(  y\right)  ~~~\frac{\mathsf{erg}%
}{\mathsf{cm}^{3}\mathsf{s}}, \label{PBF3P2}%
\end{equation}
~ with
\begin{equation}
F_{\mathrm{PBF}}\left(  y\right)  =y^{2}\int_{0}^{\infty}dx\frac{z^{4}%
}{\left(  1+\exp z\right)  ^{2}}, \label{FPBF3P2}%
\end{equation}
and%
\begin{equation}
\epsilon_{\mathrm{SWD}}=1.\,\allowbreak37\times10^{19}~\left(  \frac{M^{\ast}%
}{M}\right)  \left(  \frac{p_{F}}{Mc}\right)  T_{9}^{7}\mathcal{N}_{\nu
}C_{\mathrm{A}}^{2}\frac{\left(  y/\sqrt{5}\right)  ^{7}\mathcal{I}_{0}\left(
y\right)  }{\exp\left(  y/\sqrt{5}\right)  -1}~~~\frac{\mathsf{erg}%
}{\mathsf{cm}^{3}\mathsf{s}}. \label{ESWD}%
\end{equation}%
\[
\epsilon_{\mathrm{SWD}}=1.\,\allowbreak37\times10^{19}~\left(  \frac{M^{\ast}%
}{M}\right)  \left(  \frac{p_{F}}{Mc}\right)  T_{9}^{7}\mathcal{N}_{\nu
}C_{\mathrm{A}}^{2}\frac{\left(  y\omega_{1}/\Delta\right)  ^{7}%
\mathcal{I}_{0}\left(  y\right)  }{\exp\left(  y\omega_{1}/\Delta\right)
-1}~~~\frac{\mathsf{erg}}{\mathsf{cm}^{3}\mathsf{s}}.
\]
where%
\begin{equation}
\mathcal{I}_{0}\left(  y\right)  =\int_{0}^{\infty}\frac{du}{\left(
u^{2}+1\right)  ^{3/2}}\tanh\frac{y}{2}\sqrt{u^{2}+1}. \label{I0}%
\end{equation}
\begin{figure}[ptb]
\includegraphics{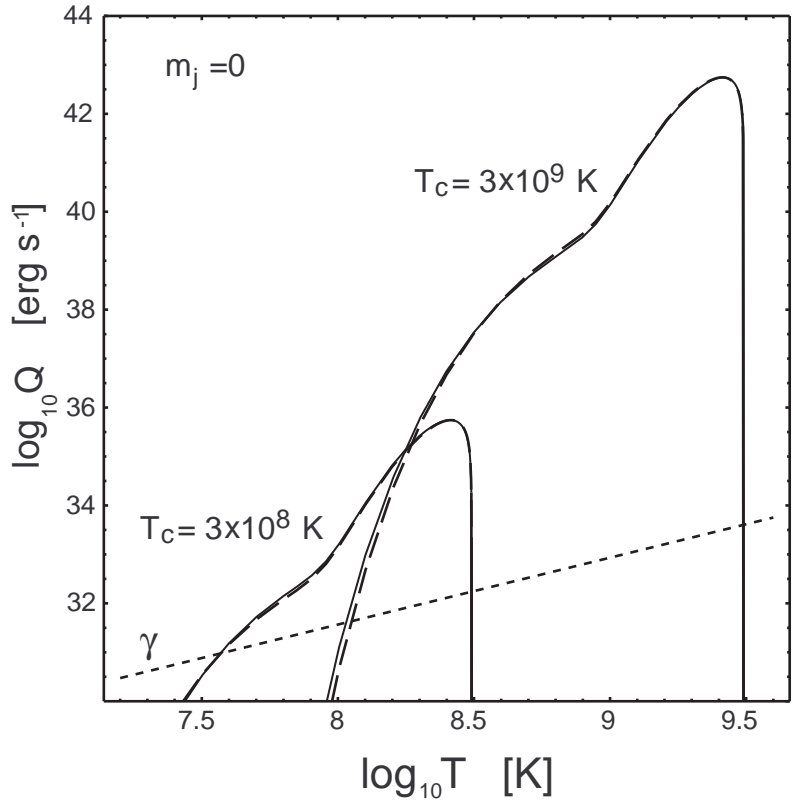}\caption{Temperature dependence of the total bulk
neutrino luminosity from a homogeneous superfluid core owing to recombination
of Cooper pairs (PBF) and owing to decay of spin waves (SWD) for
$p_{F}=2.1~fm^{-1}$. The effective mass is taken to be $M^{\ast}=0.7M$. Solid
lines are calculated according to the exact expressions (\ref{ergPBF}),
(\ref{ergSWD1}), (\ref{ergSWD2}). Dash lines are calculated by simplified
formulas (\ref{PBF3P2}) and (\ref{ESWD}). Volume of the triplet condensate is
estimated as $7\times10^{18}\ cm^{3}$. The short-dashed line is the energy
losses per unit of time owing to the surface $\gamma$-radiation, as calculated
in Ref. \cite{Page04}.}%
\label{lum}%
\end{figure}The accuracy of Eq. (\ref{ESWD}) substantially increases if one
takes into account that the spin wave energy $\omega_{1}$ depends on the
temperature. As found in Ref. \cite{L10a} this dependence can be evaluated
making use of the analytic fit which relates $\Omega_{1}$ to $y$ at any
$y>0$:
\[
\frac{\omega_{1}}{2\Delta}=\frac{0.2172-0.0059y+0.0114y^{2}+0.0026y^{3}%
}{1+0.0534y+0.0710y^{2}+0.0175y^{3}}~.
\]
The maximum fit error is about $0.1$\%.

To get an idea of the accuracy of the simplified expressions, in Fig.
\ref{lum} we demonstrate the total neutrino energy losses caused by PBF and
SWD neutrino emission from the superfluid core of the volume $7\times
10^{18}\ cm^{3}$. The total neutrino luminosity, as calculated by the exact
Eqs. (\ref{ergPBF}), (\ref{ergSWD1}), and (\ref{ergSWD2}), is shown in
comparison with the sum of the PBF and SWD neutrino losses calculated with the
aid of simple expressions given by Eqs. (\ref{PBF3P2}) and (\ref{ESWD}). We
display also the the luminosity of the surface photon radiation. The latter is
taken as in Fig. 20 of Ref. \cite{Page04}.

\section{Summary and conclusion}

According to modern theories the triplet-spin superfluid condensate in the
inner core of neutron stars arises owing to pairing of neutrons caused by
attractive spin-orbit and tensor forces and consists of the orbital
contributions corresponding to $l=1,3$. Assuming the projection of the total
angular momentum $m_{j}=0$ , the superfluid energy gap $\Delta$ arising in the
$^{3}P_{2}-$\noindent$^{3}F_{2}$ state is $\Delta^{2}=\Delta_{1}^{2}%
+\Delta_{3}^{2}$, where the contribution $\Delta_{3}\neq0$ is caused by the
tensor interactions. We have studied the influence of the admixture of the
$^{3}F_{2}$ state onto the collective spin oscillations and neutrino emission
processes in the triplet superfluid neutron liquid.

To evaluate the rate of neutrino energy losses out of the $^{3}P_{2}%
-$\noindent$^{3}F_{2}$ superfluid neutron liquid we have calculated the
anomalous three-point vertices responsible for the interaction of the
superfluid liquid with an external axial-vector field. The BCS-like
calculation has done in the angle average approximation. The polelike behavior
of the vertices points out the existence of two twofold eigen modes of
oscillations. The oscillation frequencies in terms of the gap components are
given by Eqs. (\ref{w3}) and (\ref{w4}). According to the obtained expressions
the known low-frequency mode $\omega_{1}\simeq\Delta/\sqrt{5}$ undergoes only
a small frequency shift owing to the tensor interactions. The frequency of the
new, upper mode $\omega_{2}\left(  \Delta_{3}=0\right)  \simeq\sqrt
{\allowbreak58/35}\Delta$~decreases slightly along with increasing of the
tensor contribution into the energy gap. We found that the excitation of the
high-frequency spin oscillations is strongly quenched if the tensor
interactions between the pairing particles are not taken into account, that
is, if $\Delta_{3}=0$. According to calculations of different authors, at the
Fermi surface one has $\Delta_{3}\simeq0.17\Delta_{1}$ (see, \textit{e.g.},
Ref. \cite{Khod}). In this case our theoretical analysis predicts two twofold
modes $\omega=\omega_{1}\simeq0.42\Delta$ and $\omega=\omega_{2}%
=1.\,\allowbreak19\Delta$.

We have derived the linear response of the superfluid liquid onto an external
axial-vector field. At the time-like momentum transfer the imaginary part of
this polarization function consists of three contributions originating from a
recombination of broken Cooper pairs and from weak decays of the collective
modes of spin oscillations. Accordingly, the neutrino energy losses through
neutral weak currents consist of three contributions caused by the above decay
processes. The neutrino energy losses owing to PBF, SWD$_{1}$, and SWD$_{2}$
processes are presented analytically by Eqs. (\ref{ergPBF}), (\ref{ergSWD1}),
and (\ref{ergSWD2}).

Neutrino decays of the low-energy spin waves (SWD$_{1}$) can play an important
role in the cooling scenario of neutron stars. Previously we have demonstrated
(see Fig. 5 in Ref. \cite{L10b}) that the decays of spin waves with
$\omega=\Delta/\sqrt{5}~$can become the dominant cooling mechanism in a wide
range of low temperatures and modify the cooling trajectory of neutron stars.

Weak decays of the high-frequency mode (SWD$_{2}$) occur only if the tensor
forces are taken into account in the pairing interaction, that is, if
$\Delta_{3}\neq0$. The maximal neutrino emission caused by the SWD$_{2}$
processes is of the same order as in the SWD$_{1}$, however the neutrino
energy losses from the decay of the upper mode decrease more rapidly along
with lowering of the temperature. As a result the SWD$_{2}$ contribution into
the total energy losses is negligible in comparison with the sum of the PBF
and SWD$_{1}$ contributions. This fact makes it possible to neglect the
SWD$_{2}$ contribution and describe the neutrino energy losses from the
$^{3}P_{2}-$\noindent$^{3}F_{2}$ superfluid liquid by simple expressions given
by Eqs. (\ref{PBF3P2}) and (\ref{ESWD}).


\begin{thebibliography}{99}                                                                                               %


\bibitem {YL}O. V. Maxwell, Astrophys. J. 231 (1979) 201.

\bibitem {Page04}D. Page, J. M. Lattimer, M. Prakash, A. W. Steiner,
Astrophys. J. Supp. 155 (2004) 623.

\bibitem {Page09}D. Page, J. M. Lattimer, M. Prakash, A. W. Steiner,
Astrophys. J. 707 (2009) 1131.

\bibitem {SY}P. S. Shternin, D. G. Yakovlev, C. O. Heinke, W. C. G. Ho, D. J.
Patnaude, Mon. Not. Roy. Astron. Soc. 412 (2011) L108.

\bibitem {PPLS}D. Page, M. Prakash, J. M. Lattimer, A. W. Steiner, Phys. Rev.
Lett. 106 (2011) 081101

\bibitem {Maki}K. Maki and H. Ebisawa, J.Low Temp. Phys. 15 (1974) 213.

\bibitem {C1}R. Combescot, Phys. Rev. A 10 (1974) 1700.

\bibitem {W}P. W\"{o}lfe, Phys. Rev. Lett. 37 (1976) 1279.

\bibitem {Wolfle}P. W\"{o}lfe, Physica B 90 (1977) 96.

\bibitem {L09a}L. B. Leinson, Phys. Rev. C 81, 025501 (2010).

\bibitem {L10a}L. B. Leinson, Phys. Lett. B 689 (2010) 60.

\bibitem {L10b}L. B. Leinson, Phys. Rev. C 82, 065503 (2010).

\bibitem {Tamagaki}R. Tamagaki, Prog. Theor. Phys. 44 (1970) 905.

\bibitem {Takatsuka}T. Takatsuka, Prog. Theor. Phys. 48 (1972) 1517.

\bibitem {AGD}A. A. Abrikosov, L. P. Gorkov, I. E. Dzyaloshinkski,
\textit{Methods of quantum field theory in statistical physics}, (Dover, New
York, 1975).

\bibitem {Migdal}A. B. Migdal, \textit{Theory of Finite Fermi Systems and
Applications to Atomic Nuclei} ~(Interscience, London, 1967).

\bibitem {Amundsen}L. Amundsen and E. \O stgaard, Nucl. Phys. A 437 (1985) 487.

\bibitem {Khod}V. V. Khodel, V. A. Khodel, and J. W. Clark, Nucl. Phys. A 679
(2001) 827.

\bibitem {Baldo}M. Baldo, J. Cugnon, A. Lejeune and U. Lombardo, Nucl. Phys. A
536 (1992) 349.

\bibitem {Elg}\O . Elgar{\o }y, L. Engvik, M. Hjorth-Jensen, E. Osnes, Nucl.
Phys. A 607 (1996) 425.

\bibitem {Khodel}M.V. Zverev, J. W. Clark, and V. A. Khodel, Nucl. Phys. A 720
(2003) 20.

\bibitem {Schwenk}A. Schwenk and B. Friman, Phys. Rev. Lett. 92, C82501 (2004).

\bibitem {Larkin}A. I. Larkin and A. B. Migdal, Zh. Experim. i Teor. Fiz. 44
(1963) 1703 [Sov. Phys. JETP 17 (1963) 1146].

\bibitem {Leggett}A. J. Leggett, Phys. Rev. 140 (1965) 1869.

\bibitem {Leggett1}A. J. Leggett, Phys. Rev. 147 (1966) 119.

\bibitem {TakTam}T. Takatsuka and R. Tamagaki, Prog. Theor. Phys. Suppl. 112
(1993) 27.

\bibitem {BEEHS}M. Baldo, \O . Elgar{\o }y, L. Engvik, M. Hjorth-Jensen, and
H.-J. Schulze, Phys. Rev. C 58 (1998) 1921.

\bibitem {VVkh}V. A. Khodel, V. V. Khodel, and J. W. Clark,\textbf{ }Nucl.
Phys. A 679 (2001) 827.

\bibitem {L01}L. B. Leinson, Phys. Rev. C 78, 015502 (2008).

\bibitem {LP06}L. B. Leinson and A. P\'{e}rez, Phys. Lett. B 638 114 (2006).

\bibitem {YKL}D. G. Yakovlev, A. D. Kaminker, and K. P. Levenfish, Astron.
Astrophys. 343 (1999) 650.
\end{thebibliography}
\end{document}